\documentclass[runningheads]{llncs}
\usepackage{float}
\usepackage{caption}
\usepackage{subcaption}
\usepackage{url}
\usepackage{paralist}
\usepackage{todonotes}
\usepackage[pdfpagelabels,colorlinks,allcolors=blue]{hyperref}
\usepackage{xspace}
\usepackage{algorithm}
\usepackage{algpseudocode}
\usepackage{amsfonts}
\usepackage{amsmath}
\usepackage[inline]{enumitem}
\usepackage{gensymb}

\newcommand{\sixth}{0.15\textwidth}

\usepackage{array}
\newcolumntype{C}[1]{>{\centering\let\newline\\\arraybackslash\hspace{0pt}}m{#1}}

\usepackage{graphicx}
\graphicspath{{figures/}}
\usepackage{makecell}

\begin{document}

\title{On the Perception of Small Sub-graphs}
\titlerunning{On the Perception of Small Sub-graphs}
%
\author{Jacob Miller \and Mohammad Ghoniem \and Hsiang-Yun Wu \and Helen C. Purchase\\
        \small{
            \url{jacobmiller1@arizona.edu} \and 
            \url{mohammad.ghoniem@list.lu} \and
            \url{hsiang.yun.wu@gmail.com} \and
            \url{helen.purchase@monash.edu}  
        }
    }


\institute{
    Department of Computer Science, University of Arizona \and 
    Luxembourg Institute of Science and Technology \and 
    St. Pölten University of Applied Sciences \and 
    Department of Human-Centred Computing, University of Monash 
}

\authorrunning{Miller et al.}

%
\maketitle              

\begin{abstract}
    Interpreting a node-link graph is enhanced if similar sub-graphs (or `motifs') are depicted in a similar manner -- that is, they have the same visual form. Small motifs within graphs may be perceived to be identical when they are structurally dissimilar, or may be perceived to be dissimilar when they are identical. This issue primarily relates to the Gestalt principle of similarity, but may also include an element of quick, low-level pattern-matching. We believe that if motifs are identical, they should be depicted identically; if they are nearly-identical, they should be depicted nearly-identically. This principle is particularly important in domains where motifs hold meaning and where their identification is important. 
    We identified five small motifs: bi-cliques, cliques, cycles, double-cycles, and stars. For each, we defined visual variations on two dimensions – same or different structure, same or different shape. We conducted a crowd-sourced empirical study to test the  perception of similarity of these varied motifs, and found that determining whether motifs are identical or similar is affected by both shape and structure. 
\end{abstract}

\section{Introduction}
\label{sec:intro}
\begin{figure}
    \centering
    \includegraphics[width=0.32\textwidth]{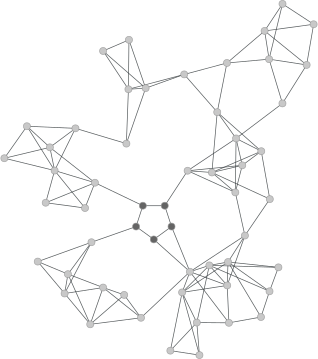}
    \hfill \includegraphics[width=0.32\textwidth,angle=90]{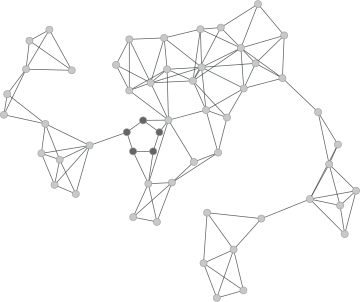}
   \hfill \includegraphics[width=0.32\textwidth,angle=90]{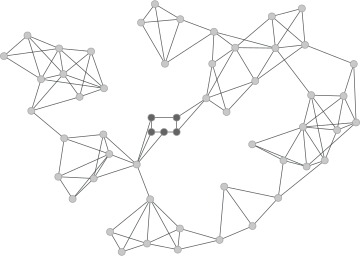}
    
    \parbox[c]{0.32\linewidth}{\centering (a) Base motif\\~}\hfill
    \parbox[c]{0.32\linewidth}{\centering (b) Same-structure/\\same-shape}\hfill
    \parbox[c]{0.32\linewidth}{\centering (c) Same-structure/\\different-shape}

    \hspace{0.10\textwidth}
    \hfill
    \includegraphics[width=0.32\textwidth]{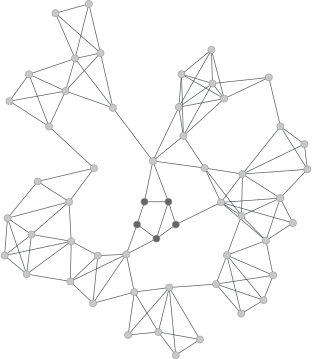}
    \hfill
    \includegraphics[width=0.32\textwidth]{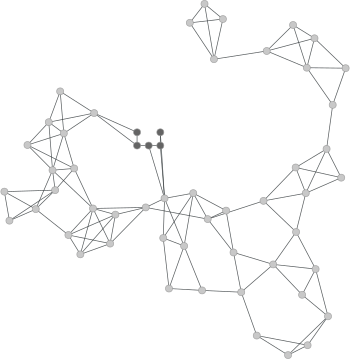}
    \hfill
    \hspace{0.10\textwidth}

    \parbox[c]{0.32\linewidth}{\centering (d) Different-structure/\\similar-shape}
    \parbox[c]{0.32\linewidth}{\centering (e) Different-structure/\\different-shape}
    
    \caption{Example `cycle' motif drawings. (a) shows the base shape and structure are regular and well-formed; nodes are placed on a pentagon. (b)--(e) show variations of shape (visual form) and/or structure (connectivity). 
    }
    \label{fig:teaser}
\end{figure}

As they attempt to understand complex natural, technical or social phenomena, application domain experts use graph models to analyze intricate relationships.
They often resort to graph visualization tools to navigate in and make sense of such data~\cite{graphvis-survey}.
Much work has also looked at the automatic extraction of graph motifs~\cite{micale2018fast}, and at counting them~\cite{JERRUM2015702}, to characterize important graph-level properties.
Graph motifs are simply local connectivity patterns, or small sub-graphs, lying within graphs; see \autoref{fig:teaser}\,(a).
They can be elementary motifs, like triangles, or assembled into higher-order motifs, like chains and cycles~\cite{PhysRevE.98.062312}.
Given their role in anticipating the behavior of the graph at hand, there is so far little work dedicated to the visual perception of graph motifs. 
In particular, existing graph visualization tools are usually not designed to ensure that similar graph motifs take a similar graphical form.

\autoref{fig:teaser} uses a five-cycle motif to exemplify the problem of visual matching of graph motifs. Motif nodes are highlighted in black, while other nodes are grey. In \autoref{fig:teaser}\,(a), the base shape places the motif nodes on the vertices of a regular pentagon. In \autoref{fig:teaser}\,(b), the same structure is also drawn as a regular pentagon, subject to a rotation. Compared to the drawing of the base shape, one might hypothesize that users will recognize that the two motifs are identical. In \autoref{fig:teaser}\,(c), the motif is drawn as a quadrilateral, with the fifth node lying within, making it harder to match it to the base shape. In \autoref{fig:teaser}\,(d), the pentagon shape is preserved despite the addition of an edge to the motif. The rationale is that a small structural change should be commensurate to the induced change of shape. In \autoref{fig:teaser}\,(e), both the structure and the shape are different to the base motif and its shape, which users should see easily. \autoref{sec:stimuli} discusses the full list of motifs considered in this study (see \autoref{fig:star}--\autoref{fig:biclique}).

In this paper, we explore the perception of (identical or similar) sub-graphs by conducting an empirical study which asks participants to compare (identical or similar) depictions of sub-graphs. Our results include evidence that depicting identical sub-graphs differently makes them harder to recognise as the same, and depicting different sub-graphs similarly results in false identically judgements.

\section{Background and Related Work}


Similarity is essential in knowledge development as it allows us to organize principles to classify, form, and generalize concepts~\cite{Tversky:PR:1977}.
It also serves as a common measure for
comparison purposes or tasks in data visualization~\cite{vonLandesberger:CGA:2018}.
Specific shapes in visualization, such as \emph{Star Glyphs}~\cite{Fuchs:TVCG:2014}, \emph{Scatterplots}~\cite{Pandey:CHI:2016}, and \emph{Directed Acyclic Graphs (DAGs)}~\cite{Ballweg:GD:2018}, have been considered as factors that influence human similarity perception.
One particular usage of similarity in graph analytics is \emph{motif analysis}, because structural motifs in graphs often 
act as influential building blocks in many domains~\cite{Stone:CB:2019}, such as biology~\cite{Redhu:BI:2022}, social science~\cite{wasserman1994social}, internet communication, and others.
Therefore, visually identifying these sub-graph structures (motifs) facilitates effective comparisons among various data.

\subsection{Perception of Similarity and Shape}

Our research questions draw on well-established research on human perception. Ware’s three levels of perceptual processing ~\cite{Ware:2021:BOOK} start with a `bottom-up' stage that processes low-level visual properties in parallel, identifying, e.g., colour, texture, movement etc.; it is quick, automatic and data-driven. The ‘pattern recognition’ second level comprises sequential processing of the scene, recognising patterns, contours, and regions. The slower third level `top-down' phase sequentially explores the scene to identify objects, typically engaging cognition for the completion of a specific task. 
The first level includes the immediate identification of prominent objects which ‘pop-out’, being of obvious different visual form to those surrounding them. The visual features that result in pop-out are of varying effectiveness: colour is the most obvious one; others include texture, orientation, size, shape, curvature~\cite{Treisman:1985:CVGI}. 
The Gestalt laws \cite{Cherry:Gestalt:2023,Koffka:PGP:1935} describe how we see patterns and groups, and relate to the second of Ware’s levels. Distinct objects may be seen to form a group by being close together (proximity), looking similar (similarity), or moving together (common fate), etc.
This paper considers the quick recognition of shapes and the Gestalt law of similarity in graph drawing: if two sub-graphs are of the same structure, then depicting them in the same visual form will ensure that they can be quickly recognised as the same; if they are of similar structure, then depicting them in the same (or similar) visual form will highlight their similarity. We investigate the immediate recognition of same (or similar) sub-graphs rather than serial processing requiring the use of cognition because making similarities immediately prominent can help in gaining a better (and quicker) overall understanding of the structure of a graph at a glance. Comparing sub-graphs with enough time to engage cognitive processes is simple; making such comparisons quickly may not be. Our research questions focus on determining whether depicting identical sub-graphs using the same (or similar) shape facilitates the recognition of their identicality. We also explore what happens when the sub-graph structure is slightly different, and when its presentation is distorted by additional forces within a force-directed algorithm.

\subsection{Related Work}
\label{ssec:related}

Prior work has investigated the perception of graph properties, such as graph density, clustering coefficient~\cite{Soni:2018:CGF} and layout types~\cite{Kypridemou:2022:DRI}. Although the perception of shapes in sub-graphs {\it per se} has not yet been fully investigated with respect to sub-graph structure, some studies of shapes in visualization have been conducted. 
Gogolou et al. investigated if time series visualizations generated from automatic similarity measures are aligned with readers' similarity constraints~\cite{Gogolou:TVCG:2019}.
They concluded that the selection of visualizations influences the patterns that readers consider as similar. 
Ballweg et al. researched the influencing factors of directed acyclic graphs (DAGs) when they are drawn as layered drawings~\cite{Ballweg:GD:2018}. Their study suggests that the similarity perception of DAGs is mainly affected by the number of levels in the drawings, the number of nodes on a level, and the corresponding overall shape (i.e., convex hull of the DAG), while Wallner et al.~\cite{Wallner:2019:CE,Wallner:2020:DRI} did not find significance in the perception of overall shapes.
Tarr and Pinker studied the effect of mental rotation in shape recognition, focusing on letter-like asymmetrical characters~\cite{Tarr:CP:1989}. 
They found that trained readers can recognize the characters almost as fast at all  familiar angles, while the performance varies when the character appears at novel angles.

Besides perceptual studies, several graph drawing algorithms incorporate shapes (i.e., motifs from different graph classes~\cite{deRidder:ISGCI:2014}) to improve layout readability.
Yuan et~al. developed an algorithm based on Laplacian constrained distance embedding to control the shapes of sub-graphs input by the users~\cite{Yuan:VIS:2012}.
Wang et~al. generalized the classical stress majorization approach~\cite{Wang:VIS:2018}, to ease sub-graph shape manipulation.
Meidiana et al. extended the classical force-directed and stress minimization algorithms to optimize shape-based metrics measure~\cite{Eades:GD:2015} to achieve faithful drawings~\cite{Meidiana:GD:2023}.
Interactions through aggregation~\cite{Landesberger:2009:VMV} and simplification~\cite{Cody:2013:CHI} also consider motif properties to reduce layout visual complexity.

In practice, many applications consider the visual form of motifs, including transcriptional regulation networks analysis~\cite{Huang:2005:IV}, phylogenetic trees comparison~\cite{Bremm:2011:VAST}, biological pathways diagrams~\cite{Yuan:VIS:2012}, metro maps creation~\cite{Batik:PG:2022,Wu:CGF:2020}, dynamic graph analysis~\cite{Cakmak:VDS:2022}, and visual graph matching~\cite{graph-matching}, for example, across distinct layers of a multilayer graph~\cite{multilayer}.
In this paper, we demonstrate the benefit of depicting motifs within graphs in a consistent, well-formed manner.


\section{Methodology}
\label{sec:method}

\subsection{Stimuli}
\label{sec:stimuli}

\begin{figure}
\centering
    \includegraphics[width=0.19\textwidth]{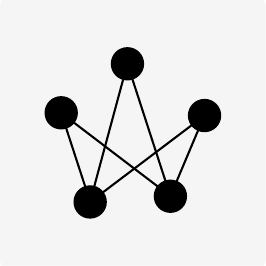}
    \includegraphics[width=0.19\textwidth]{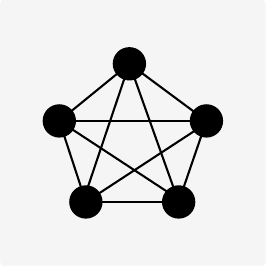}
    \includegraphics[width=0.19\textwidth]{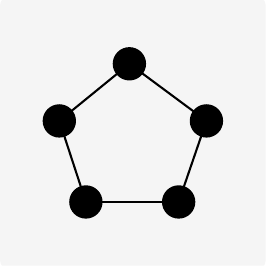}
    \includegraphics[width=0.19\textwidth]{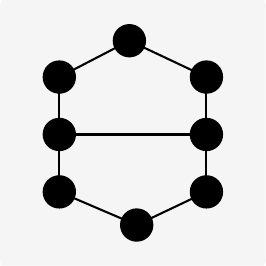}
    \includegraphics[width=0.19\textwidth]{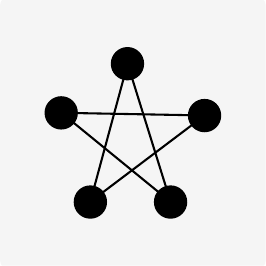}

    \parbox[c]{0.19\linewidth}{\centering (a) bi-clique}
    \parbox[c]{0.19\linewidth}{\centering (b) clique}
    \parbox[c]{0.19\linewidth}{\centering (c) cycle}
    \parbox[c]{0.19\linewidth}{\centering (d) double-cycle}
    \parbox[c]{0.19\linewidth}{\centering (e) star}    

\caption{The five base motifs.}
\label{fig:stimuli}
\end{figure}

Structural motifs often bind with semantics in applications for analysis purposes (\autoref{ssec:related})~\cite{PhysRevE.98.062312}. For example, a clique (or a bi-clique) in social networks represents strong mutual connections~\cite{wasserman1994social}, and a cycle 
in biological networks indicates a circular biochemical reaction~\cite{Leontis:2006:COSB,Royer:CB:2008}. 
We hence identified five `motifs': small sub-graphs with well-defined graph structure, distinct visual form, easy description and clear definition. We denote these motifs (the graph structure along with node positions) as our `base' motifs; see \autoref{fig:stimuli}. 
For each base motif, we created 12 variations according to two dimensions: change of structure (Yes/No) and change of shape (Yes/No), where shape is determined by the convex hull of nodes (smallest convex set containing all nodes). There are hence four quadrants. 

\begin{figure}[h!]
\centering
    \includegraphics[width=0.9\linewidth]{"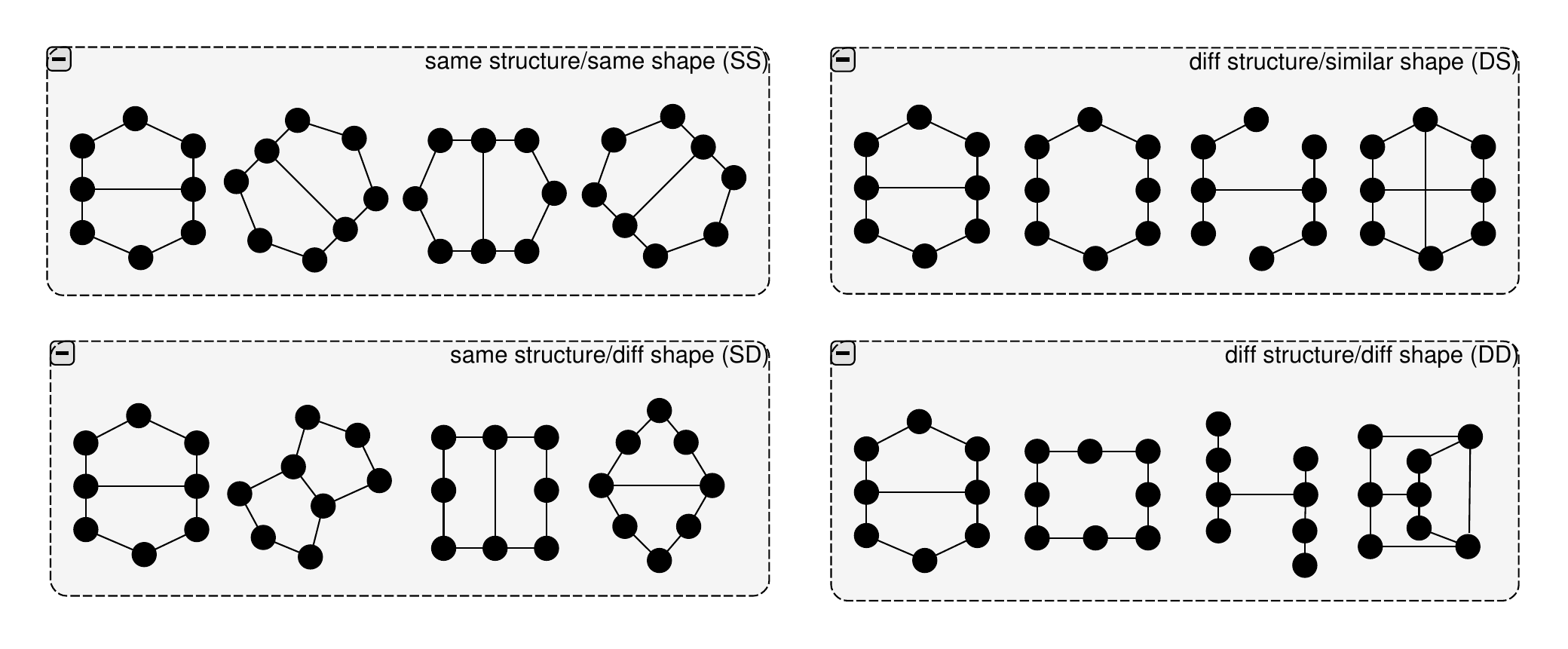"}
\caption{12 variations in shape and/or structure of the \emph{double cycle} motif.}
\label{fig:dcycle}
\centering
    \includegraphics[width=0.9\linewidth]{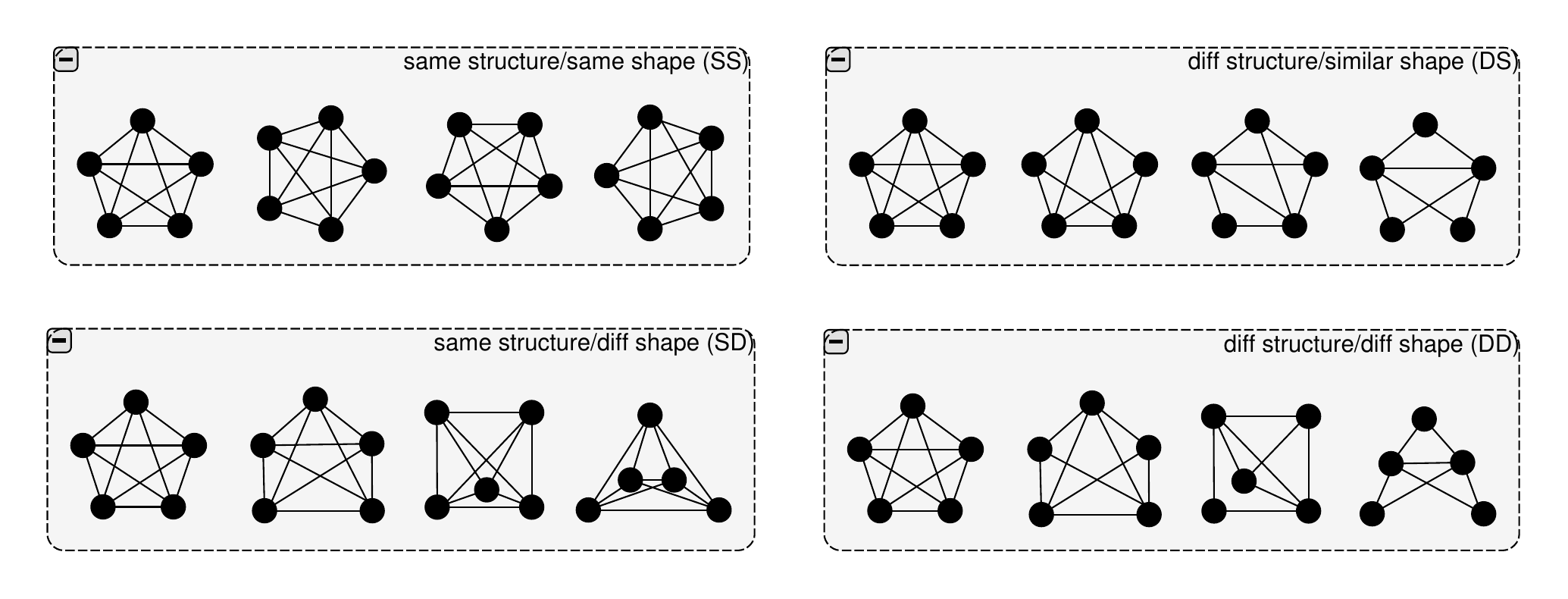}
\caption{12 variations in shape and/or structure of the \emph{clique} motif.}
\label{fig:clique}
\centering
    \includegraphics[width=0.9\linewidth]{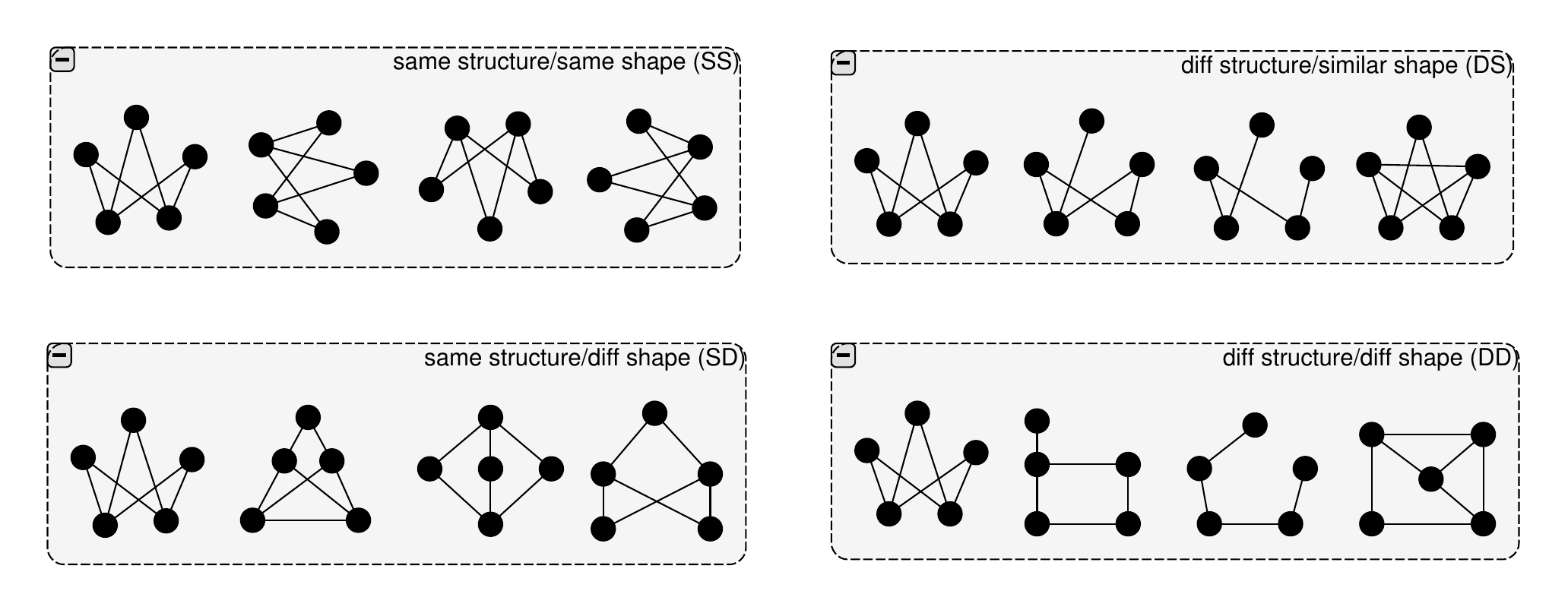}
\caption{12 variations in shape and/or structure of the \emph{bi-clique} motif.}
\label{fig:biclique}
\end{figure}

The \textbf{same-structure/same-shape (SS)} variations depict the same graph structure and same visual form as the base, rotated by 90, 180, 270$\degree$ (45, 90 and 135$\degree$ for double-cycles which have two axes of symmetry (\autoref{fig:dcycle}\ (top left)). 

\begin{figure}[htbp]
\centering
    \includegraphics[width=0.9\linewidth]{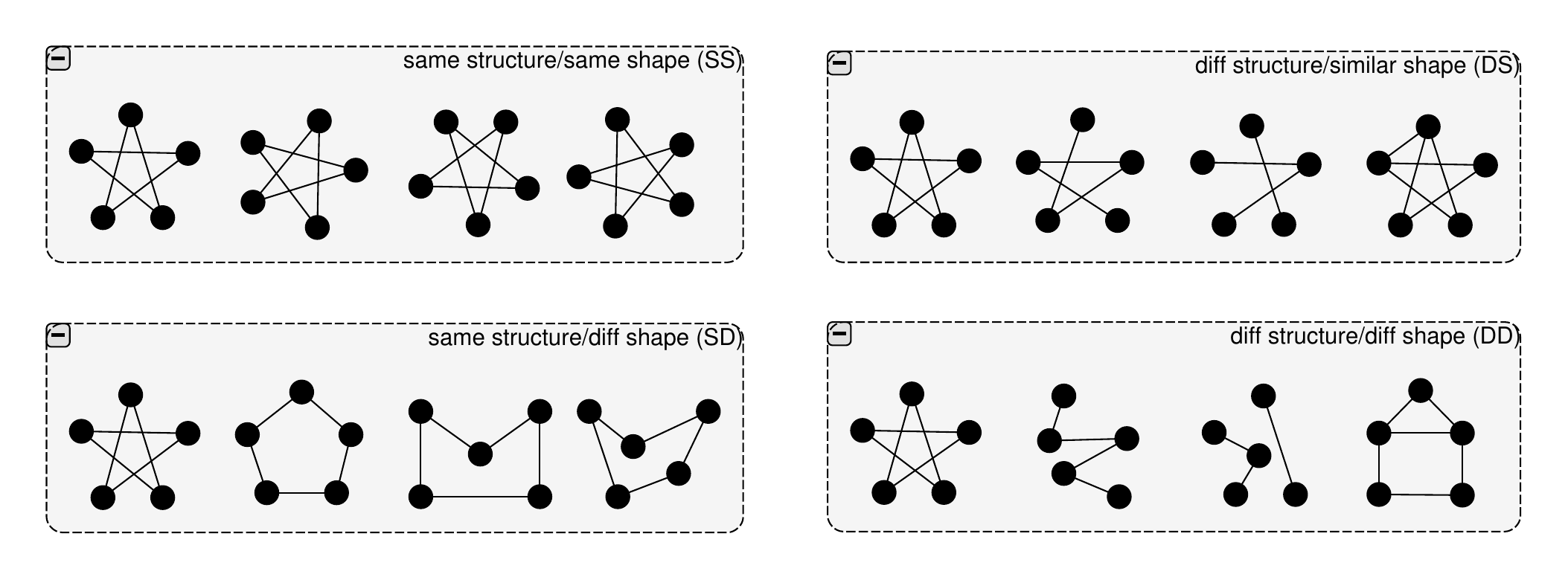}
\caption{12 variations in shape and/or structure of the \emph{star} motif.}
\label{fig:star}
\centering
    \includegraphics[width=0.9\linewidth]{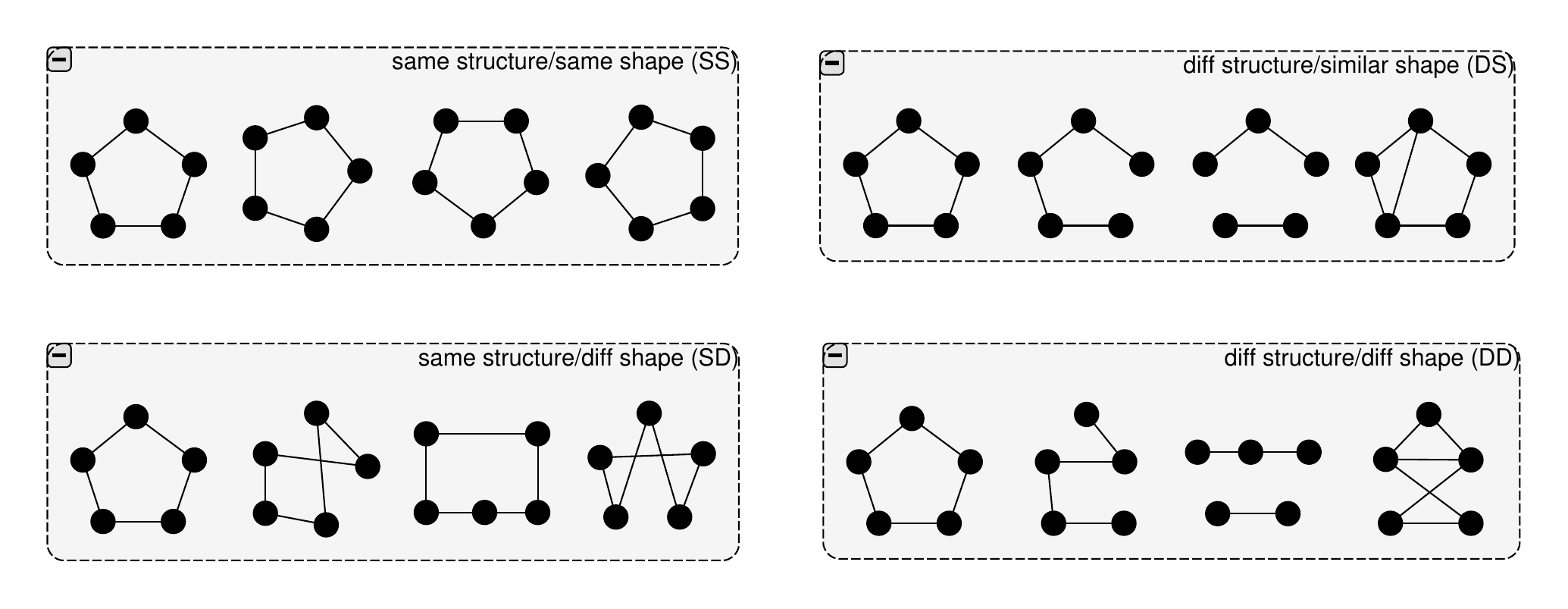}
\caption{12 variations in shape and/or structure of the \emph{cycle} motif.}
\label{fig:cycle}
\end{figure}

The \textbf{same-structure/different-shape (SD)} variations use the same graph structure as the base motif but a different shape; we generated three different drawings by varying the relative positions of nodes (e.g.~\autoref{fig:star} (bottom left)). 

The \textbf{different-structure/similar-shape (DS)} variants adopt a different graph structure to the motif; we change the number of edges (deleting one, deleting two, or adding one) while keeping node positions as in the base -- giving them a similar (but not identical) shape (e.g.~\autoref{fig:star}\ (top right)). As edges can not be added to the clique motif, we remove three edges (\autoref{fig:clique}\ (top right)).

The final \textbf{different-structure/different-shape (DD)} collection adapts the original motif sub-graph as for the DS variants, but depicts them using different node positions from the base motif. The three DD variants of the base motif differed both by structure and visual form (e.g.~\autoref{fig:star}\ (bottom right)).

The motif variations are: \emph{bi-clique}\,(\autoref{fig:stimuli}\,(a)), \emph{clique}\,(\autoref{fig:stimuli}\,(b)), \emph{cycle} (\autoref{fig:stimuli} (c)), \emph{double-cycle}\,(\autoref{fig:stimuli}\,(d)), and \emph{star}\,(\autoref{fig:stimuli}\,(e)).

Each motif set therefore comprises $13$ different visual motif drawings which were integrated into graphs, a total of $65$ graph drawings. The graphs and graph drawings were subject to both structural and layout constraints (\autoref{ssec:implementation}). Motif nodes were coloured black, all others grey.

\subsection{Experimental Design}
\label{ssec:design}
For each motif, we explored participants' ability to assess the identicality or similarity of motifs in two graph drawings, with each participant working with only one motif. In Experiment 1, we explored the effect of motif shape; in Experiment 2, we explored the effect of motif shape distortion. Although the $48$ trials for Experiment 1 were interspersed with the $40$ trials for Experiment 2, since their aims and stimuli were different, this first section focuses on Experiment 1.

\begin{figure}[t]
    \centering
    \fbox{\includegraphics[width=0.95\textwidth]{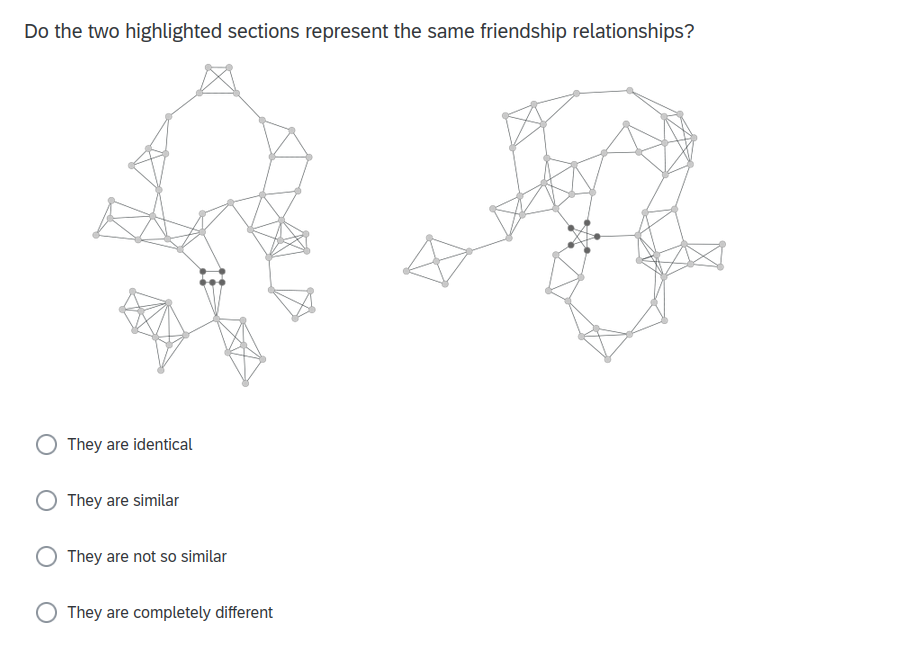}}
    \caption{Example trial -- star:  same-structure/different-shape (left); base (right).}
    \label{fig:qualtrics-examp}
\end{figure}

Each trial consisted of a pair of graph drawings displayed side-by-side, with each drawing containing a highlighted motif. Participants were asked to indicate the degree of similarity between the structure of the two motifs. For each motif, the base motif graph drawing (or one of its rotations) was paired with all other 12 motif variant drawings, giving a total of 48 trials. The position of the two drawings in each trial (left or right) was randomly determined. \autoref{fig:qualtrics-examp} shows the case of a rotated star motif on the right, with the base star motif on the left. 

Since we are interested in bottom-up immediate processing, and how known shapes can be identified quickly, we limited the length of time allocated for each trial to 4 seconds -- a duration determined by pilot testing of the complete experiment with 15 participants which considered 3, 4, and 5 seconds as possibilities: 4 seconds produced sufficient variability in accuracy to avoid floor (too hard) and ceiling (too easy) effects. Participants judged similarity on a scale by choosing among: `\emph{identical}', `\emph{similar}', `\emph{not so similar}' and `\emph{completely different}'.

\subsection{Experimental Procedure}

We used Prolific~\cite{prolific:2014} to recruit participants for our study and Qualtrics~\cite{qualtrics:2005} to implement the survey. The simplicity of our experiment makes it suited for crowd-sourcing~\cite{Borgo:2017:EC}, enabling collection from a large number of participants per motif. Participants were paid \pounds 9 per hour, with a median completion time of 10.4 minutes. They were recruited from the UK and required to use a desktop machine (no mobile devices). We ran five within-subject studies, one for each motif, and 30 participants per motif. 46\% identified as women, 52\% as men, 1\% as non-binary/gender diverse, 1\% declining to answer. 38\% were 18-35, 28\% were 36-45, 15\% were 46-55, and 19\% over the age of 55. 

Participants were shown six `practice' stimuli pairs for which data was not collected. These helped mitigate against the learning effect that can affect within-subject studies, as did the randomization of trials for each participant.
The data from participants with response rate less than 85\% were discarded, and missing responses from the remaining participants were not included in the calculation of mean accuracy data. Each participant saw each trial exactly once.

\subsection{Implementation}
\label{ssec:implementation}
\noindent 
\textbf{Graph generation and motif integration }
The motifs were integrated into larger 50-node graphs, generated using a k-nearest neighbor model with k=3 \cite{Dong:2011:WWW,Lenhof:1995:CGA}. Each graph motif was `stitched' into the larger graph by connecting each motif node to up to three non-motif nodes, chosen uniformly at random (though we modify the graph post-layout). Since the drawings focus on the highlighted motifs, the specific properties of the overall graphs (beyond their size) is not relevant.
A different graph was created for each of the 13 stimuli, for each motif.
The constraints on the graph generation were:
\begin{enumerate*}[label={(\bfseries GC\arabic*)}]
    \item All motif nodes must have at least one edge to a non-motif node.
    \item The graph must be connected. 
\end{enumerate*}


\subsubsection{Graph drawing}
We use multidimensional scaling (MDS) for graph layout \cite{Gansner:GD:2005,Zheng:TVCG:2018} for its ease of use in adding constraints and reliability in representing geometric structures.
The constraints on the creation of the graph drawings were:
\begin{enumerate*}[label={(\bfseries GD\arabic*)}]
    \item Motif nodes are fixed on the plane, so they keep the desired visual form, with the rest of the graph laid out around them. This is the basis of our research.
    \item No motif node may be on the periphery.
    \item Edges connecting non-motif nodes must not intersect edges within the motif, since this would change the visual form of the motif and will hamper its recognition in an uncontrolled manner.
\end{enumerate*}

We modify the MDS algorithm with an extra parameter $F$, standing for the  set of motif nodes whose position is fixed: 
$F = \{v | v $ is a node in a motif\,$\}$. Each member of $F$ is `pinned' in the plane so that their positions will not be updated, but the surrounding graph drawing will still be based on graph-theoretic distances. More formally, given a graph-theoretic distance matrix $d$, and a set of nodes with fixed position $F$, we want to find 
\begin{equation}
    \label{eq:opt-func}
    \min_{X_1,...X_n} \sum_{i \notin F} \sum_{j \notin F} (||X_i - X_j || - d_{i,j})^2
\end{equation}

If a drawing produced by \autoref{eq:opt-func} does not meet these requirements, we attempt to modify the drawing to fit. Since the constraints are not always simultaneously satisfiable, when they conflict we discard that drawing and try again with a new random graph. To satisfy \textbf{GD2}, we compute the convex hull of the layout, and if any motif node lies on the hull, we try again. 

\textbf{GC1} and \textbf{GD3} often conflict. To resolve, we first look at all possible $O(|V|^2)$ line segments and categorize them as \textbf{\emph{invalid}} if they violate \textbf{GD3}, i.e., if they intersect edges within the motif, or as \textbf{\emph{valid}} otherwise.
Then, we count and remove any \textbf{\emph{invalid}} non-motif edge (an edge with an endpoint not in a motif). If some motif nodes are left without non-motif edges, we satisfy \textbf{GC1} by adding to each such motif node the edge that is \textbf{\emph{valid}} and whose endpoint is nearest. Clearly, \textbf{GC1} is satisfied as we have added edges to each node which did not satisfy it, and we can verify that we did not violate \textbf{GD3} in the process.

The removal of \textbf{\emph{invalid}} edges might yield a graph that is too sparse to be realistic. 
We add as many \textbf{\emph{valid}} edges as the \textbf{\emph{invalid}} edges that were removed, shortest length first.
Finally, we check if \textbf{GC2} is met; if not, we iterate until all constraints are met.
All code is available at {\small{\url{https://github.com/Mickey253/graph-rep-sym}}}.

\section{Results}
\label{sec:results}

\subsection{Dependent variable}

 Each trial asked the participant whether the motifs were \textit{identical}, \textit{similar}, \textit{not so similar} or \textit{completely different}. In the presented social media context, we asked: ``\emph{Do the two highlighted sections represent the same friendship relationships?}''.

Measuring accuracy as a binary result (the structure is identical or not) removes subjectivity from the response, but ignores the fact that some structures are more different than others. We therefore use an accuracy measure based on graph edit distance (GED)~\cite{sanfeliu:GED:1983}. 
We determine the GED between a pair of sub-graphs: identical (GED=0); one edge difference (GED=1); two edges difference (GED=2); three edges difference (GED=3). For each of these GEDs, we score the participants' accuracy responses as shown in \autoref{tab:accuracy}.

\begin{table}[t]
    \centering
    \bgroup
    \def\arraystretch{1}
    \begin{tabular}{|c|c|c|c|c|}
        \hline
        & GED=0 & GED=1 & GED=2 & GED=3\\
        \hline
        identical &  1 & 0.6 & 0.3 & 0\\
        \hline
        similar &  0.6 & 1 & 0.6 & 0.3\\        
        \hline
        not so similar &  0.3 & 0.6 & 1 & 0.6\\ 
        \hline
        completely different &  0 & 0.3 & 0.6 & 1\\
        \hline
    \end{tabular}
    \egroup
    \vspace{0.25cm}
    \caption{Participant responses (rows) mapped to graph edit distance (columns).
    If participant response was \textit{similar} for GED=2, accuracy is recorded as 0.6.}
    \label{tab:accuracy}
\end{table}

\subsubsection{Experiment 1: Fixed Shape Motifs}
\label{sec:fixed-shape}
The pairs of graph drawing stimuli presented to the participants included the base motif in one drawing, paired with each of the 12 other stimuli. For each of the five motifs, we address the questions: 
\begin{itemize}[label={}]
    \item \textbf{Q1}: Does rotation affect the perception of the same sub-graph represented using the same visual form? 
\end{itemize}
We consider the \textit{same-structure/same-shape} variations where the base motif is compared to three of its own rotations. We expect that there will be no difference in accuracy. A repeated measure ANOVA test (\autoref{tab:q1}\,(a)) shows no significant differences, suggesting that the participants were able to correctly recognise the same motif, presented in the same-shape, regardless of its orientation. This result accords with that of Tarr and Pinker who studied the effect of mental rotation in shape recognition, focusing on letter-like asymmetrical characters~\cite{Tarr:CP:1989}.  

\begin{table}[t]
    \centering
    \begin{tabular}{|C{\sixth} | C{\sixth} C{\sixth} C{\sixth} C{\sixth} C{\sixth}|}
        \hline
        \multicolumn{6}{|l|}{
        \parbox[c]{12cm}{\textbf{(a) Q1: Effect of rotation on determining whether two motifs are identical}}} \\
        \hline 
         &  bi-clique & clique & cycle & double-cycle & star\\
         \hline
        F (df=2) & 0.476 & 0.783 & 0.492 & 0.129 & 0.146\\
        \hline 
        p & 0.623 & 0.459 & 0.613 & 0.879 & 0.865\\
        \hline \hline
        \multicolumn{6}{|l|}{ 
        \parbox[c]{12cm}{\textbf{(b) Q2: Effect of depicting different structure with similar or different shapes}}}\\
        \hline
         &  bi-clique & clique & cycle & double-cycle & star \\
         \hline
        t (df=29) & 3.289 &	2.168  &	4.230  &	0.923 &	3.780\\
        \hline 
        p (1-tailed) & 0.001	 & 0.019 &	$<$0.001	 & 0.181	& $<$0.001\\
        \hline \hline 
        \multicolumn{6}{|l|}{
        \parbox[c]{12cm}{\textbf{(c) Q3: Effect of using different shapes for the same sub-graphs}}}\\
        \hline
         &  bi-clique & clique & cycle  & double-cycle  & star \\
         \hline
        t (df=29) & 11.071	& 11.215	& 13.360 & 	7.807 & 	19.059\\
        \hline 
        p (1-tailed) & $<$0.001	 & $<$0.001 &	$<$0.001	 & $<$0.001	& $<$0.001\\        
        \hline \hline 
        \multicolumn{6}{|l|}{
        \parbox[c]{12cm}{\textbf{(d) Q4: Effect of different edit differences in different sub-graph structures. The trend line shows accuracy from no edit distance (left) to maximum edit distance (right); the accuracy scales are different for each motif}}}  \\
        \hline
         &  bi-clique  & clique  & cycle  & double-cycle  & star \\
         \hline
        F (df=3) & 11.817	& 19.719	& 3.07	& 28.850	& 36.84\\
        \hline 
        p & $<$0.001	 & $<$0.001 &	0.031 & $<$0.001	& $<$0.001\\
        \hline
         \hline  Trend line& \parbox[c]{0.15\textwidth}{
      \includegraphics[width=0.15\textwidth]{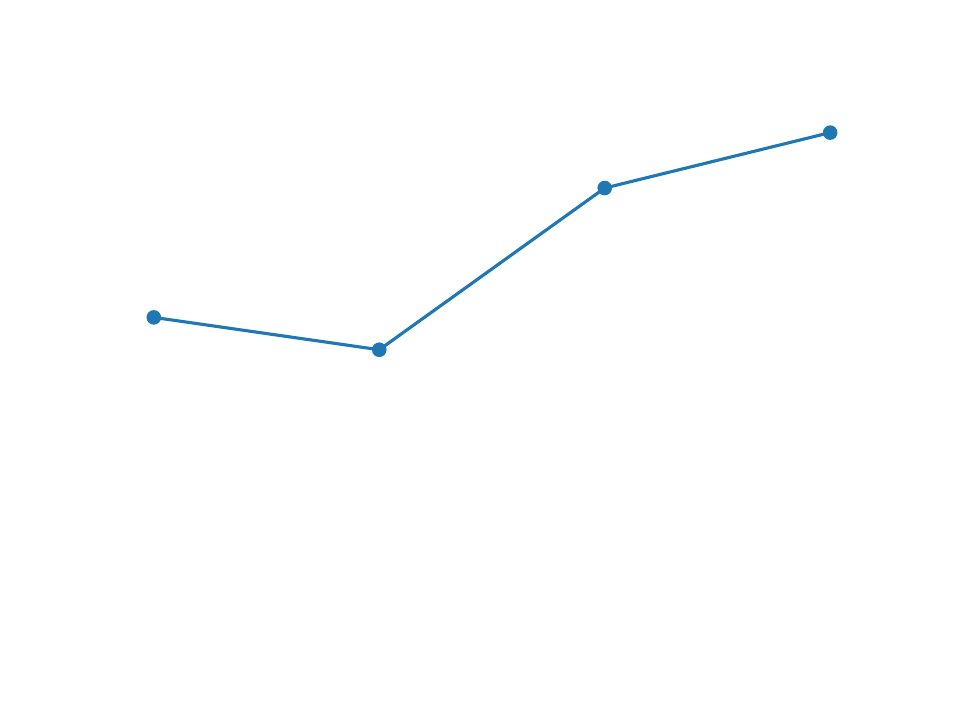}}
      & \parbox[c]{0.15\textwidth}{
      \includegraphics[width=0.15\textwidth]{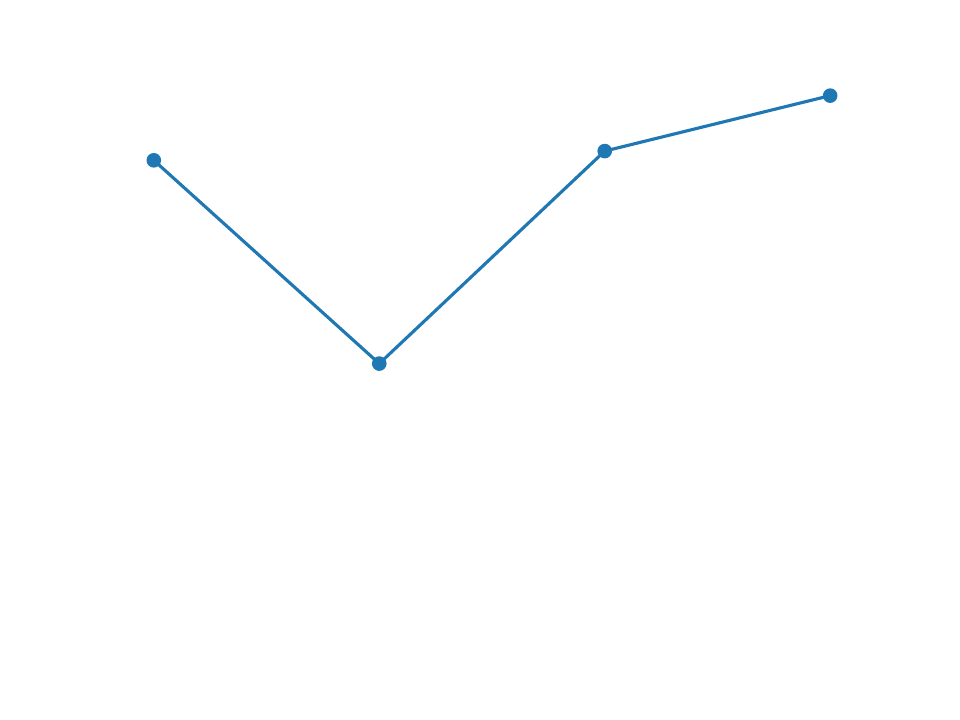}} 
      & \parbox[c]{0.15\textwidth}{
      \includegraphics[width=0.15\textwidth]{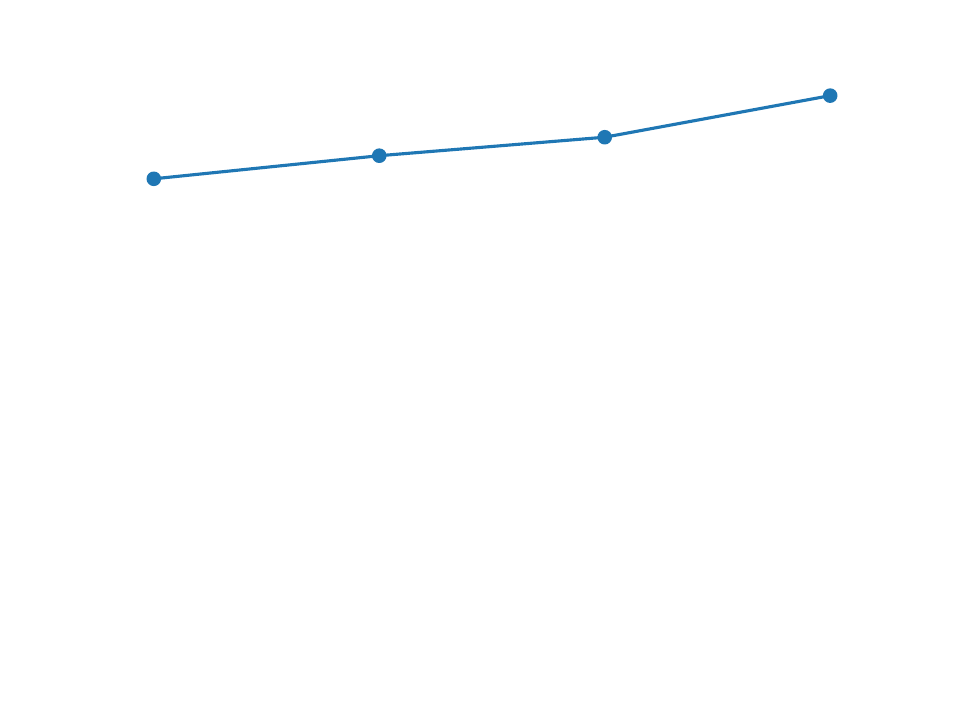}} 
      & \parbox[c]{0.15\textwidth}{
      \includegraphics[width=0.15\textwidth]{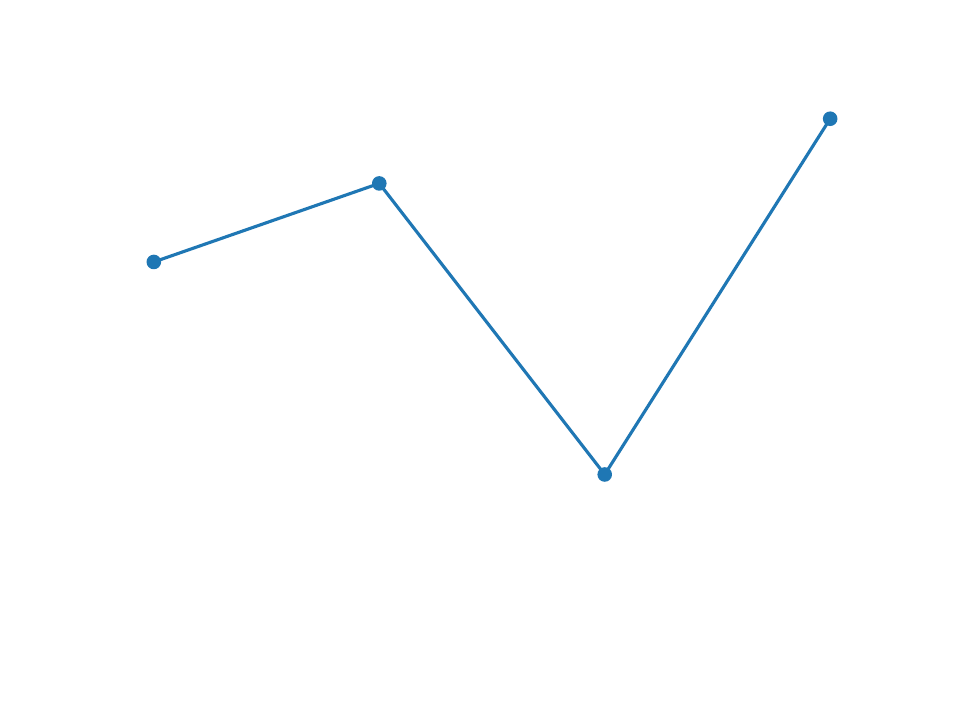}} 
      & \parbox[c]{0.15\textwidth}{
      \includegraphics[width=0.15\textwidth]{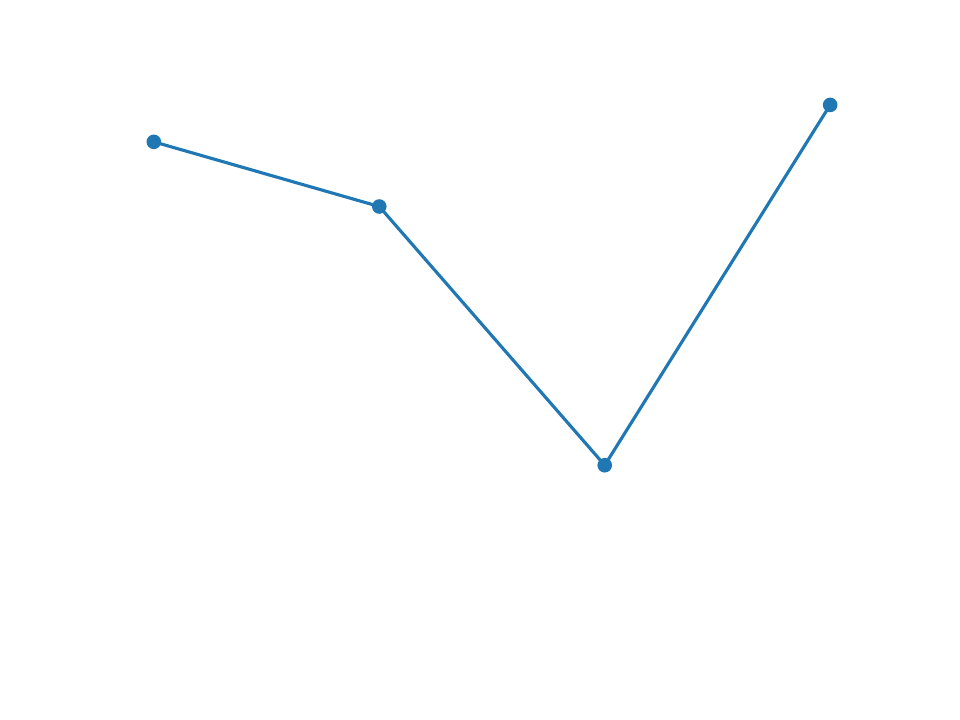}} 
      \\
      \hline 

    \end{tabular}
    \vspace{0.25cm}
    \caption{Summary of results for Experiment 1.}
    \label{tab:q1}
\end{table}

\begin{itemize}[label={}]
    \item \textbf{Q2}: If the motif is adapted to create a sub-graph of different structure, does depicting it in a similar manner affect the ability to distinguish the difference (or would it be better to depict it using a clearly different visual form?)
\end{itemize}
We compare the results of \textit{different-structure/similar-shape} with \textit{different-struc-ture/different-shape}, expecting that using a similar shape will help in identifying structural differences in the sub-graphs. A repeated measures t-test (\autoref{tab:q1}\,(b)) showed that the accuracy for \textit{different-structure/similar-shape} was significantly better than  \textit{different-structure/different-shape} for all motifs, except double-cycle.


\begin{itemize}[label={}]
    \item \textbf{Q3}: Does using a different layout affect perception of motifs of the same structure? If the motifs are identical, does it matter if they are depicted using different visual form -- or should they be depicted in the same form?
\end{itemize}
We compare  \textit{same-structure/different-shape} and \textit{same-structure/same-shape}, expecting that using the same shape for the same structure will produce better accuracy than using a different shape for identical structures. A repeated measures t-test (\autoref{tab:q1}\,(c)) showed that accuracy was significantly better for \textit{same-structure/same-shape} than for \textit{same-structure/different-shape}, for all motifs.


\begin{itemize}[label={}]
    \item \textbf{Q4}: If different sub-graphs are depicted in similar shape, are near-similar sub-graphs incorrectly assessed as being similar? 
\end{itemize}
That is, does the edit-distance between pairs of sub-graphs depicted in similar visual form affect accuracy when determining whether they are identical or not? 

Here, we focus on the three different variations in the \textit{different-structure/simi-lar-shape stimuli}, each of which has a different GED with respect to the base motif. We also include one of the \textit{same-structure/same-shape comparisons}, where GED=0. We expect that where the edit distances are small, the participants were more likely to misjudge two different sub-graphs depicted similarly as identical. A repeated measures ANOVA (\autoref{tab:q1}\,(d)) confirmed that there was an effect of edit distance, for all motifs, where accuracy generally increased with edit distance, with the greatest edit distance accuracy being at least that of the comparisons where GED=0 (those where the base motif was rotated). 

This result confirms our suggestion that the comparisons were made spontaneously – that is, in parallel, rather than serially. If they were serial, participants would have compared the motifs edge-by-edge, and accuracy for the smallest edit distances would have been higher. The spontaneous, parallel pattern matching means that stimuli of the most similar structure are incorrectly seen as identical. 

There are a few noticeable data points in this trend analysis: for double-cycle and star, adding two edges leads to lower accuracy -- both these extended motifs include at least one edge crossing. Yet, this conclusion does not extend to the bi-clique motif, although we note that the starting point of the bi-clique trend (when GED=0) is already a low accuracy score.


\begin{figure}[t]
    \centering
    \hfill    \includegraphics[width=0.35\textwidth,angle=90]{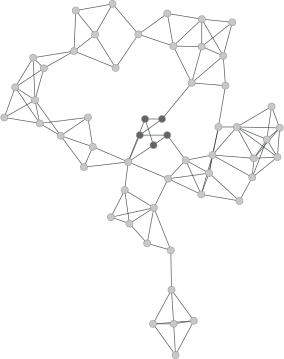}
   \hfill \includegraphics[width=0.35\textwidth,angle=90]{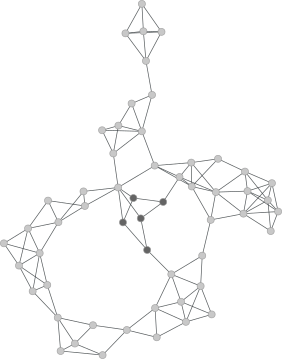}
   \hfill
    \caption{(left) Cycle motif \textit{same-structure/different-shape} variation introducing unnecessary edge crossings; (right) the same graph drawn using MDS.}
    \label{fig:cycle-examp}
\end{figure}

\subsection{Experiment 2: Unconstrained Layout}

Here, we used stimuli where the motif nodes positions were `fixed', but were distorted by MDS forces applied by surrounding nodes. For all  Experiment 1 stimuli, we created a paired `flexible' alternative, removing the three GD constraints of \autoref{ssec:implementation}, using the unmodified algorithm from~\cite{Zheng:TVCG:2018} (\autoref{fig:cycle-examp}).

Our trials comprised the `fixed' version of a motif paired with all variants of the `flexible' version, and participants were asked to judge similarity (as in Experiment 1). In each of the four (shape $\times$ structure) quadrants (\autoref{sec:stimuli}), we pair the three variants with the drawings of the same graphs rendered by an unconstrained layout, resulting in 36 pairs (4 $\times$ 3 variants (fixed) $\times$ 3 variants (flexible)). We also include the four base motif orientations: a total of 40 pairs for each motif. 
For each motif, our research question is:
\begin{itemize}[label={}]
    \item \textbf{Q5}: Is it easier to match identical small motifs if they have both been given a well-formed, regular shape (as opposed to one of them having their shape distorted by forces applied to nodes in the rest of the graph)? 
\end{itemize}
We focus on same-structure stimuli, since the question relates to identical structures, considering the shape dimension. The \textit{same-structure/same-shape} stimuli are rotations of the base; the \textit{same-structure/different-shape} stimuli, while using a different shape, are still regular and well-formed. We expect that the accuracy in matching two identical motifs with well-formed shapes will be higher than for the flexible condition (when the motifs are visually distorted). A repeated measures t-test (see \autoref{tab:q5}) showed that the accuracy when comparing two motifs with well-formed, regular shape was always significantly better than when one of the motifs had a distorted form, except in the case of the cycle. We note that the alternative well-formed, regular form of the cycle was one where two versions included unnecessary edge crossings; see \autoref{fig:cycle-examp}.

\begin{table}[t]
    \centering
    \bgroup
    \def\arraystretch{1.2}
    \begin{tabular}{|c|c c c c c|}
    \hline 
    \multicolumn{6}{|c|}{\textbf{same-structure/same-shape}}\\
    \hline
     & bi-clique  & clique & cycle  & double-cycle & star \\
    \hline
    t (df=29) & 8.692 & 11.611 & 7.380 & 10.485 & 22.254\\
    p (one-tailed) & $<$0.001  & $<$0.001  & $<$0.001  & $<$0.001  & $<$0.001 \\
    \hline 
    \multicolumn{6}{|c|}{\textbf{same-structure/different-shape}} \\
    \hline 
    t (df=29) & 6.649 & 2.582 & 1.658 & 5.477 & 8.847\\
    p (one-tailed) & $<$0.001  & 0.007 & \textbf{0.054} & $<$0.001  & $<$0.001\\
    \hline 
    \end{tabular}
    \egroup
    \vspace{0.25cm}
    \caption{The effect of using well-formed shapes to match sub-graphs.}
    \label{tab:q5}
\end{table}

\section{Discussion}

Our research questions addressed how visual form (shape) affects the interpretation of the abstract structure of a graph. We showed that by depicting two similar sub-graphs similarly, participants achieved high accuracy for all motifs but the double-cycle (\textbf{Q2)}. 
The double-cycle is the largest motif of the five, and our shape variations for it are all roughly two polygons sharing a segment. One possible cause for this is that participants were able to mentally transform the base motif (a mirrored pentagon) to the other shapes (triangle, rectangle). As a result, to gain the benefit of representing similar structures as similar shapes, it may be enough that there is a simple transformation between them, e.g., rotation or translation. 

The clear recommendation for visualization designers is that when supporting a task to identify sub-graphs one should ensure that identical sub-graphs are drawn similarly.
We have shown that rotation does not have any effect (see \textbf{Q1}), so this is a degree of freedom in visualization design, e.g., rotation of fixed motifs can be done to reduce edge crossings while maintaining shape. However, the shape of a sub-graph \textit{does} have an effect; depicting identical sub-graphs using different shapes hampers the recognition of similarity (\textbf{Q3)}. This means that even if two sections of a node-link diagram represent the same relationships, if they are drawn completely differently, one might never notice. When supporting this type of task, shape cannot be overlooked.
Designers can also remember that evaluation of similarity is done quickly (see \textbf{Q4}), so should be careful not to depict dissimilar structures using the same shape or one might wrongly interpret the data. The results of both \textbf{Q3} and \textbf{Q4} seem to indicate that a viewer will interpret similar shapes as having similar structures, regardless of ground truth.

The results of \textbf{Q3} and \textbf{Q4} can apply beyond the simple five-node motifs we have demonstrated. For instance, graphs with large defined clusters are often depicted with each cluster being drawn as large roughly circular dense regions. If the shape of these regions is roughly the same, a viewer is likely to interpret these large structures as being equivalent. If within-cluster structure is very different between clusters, a designer would want to make this apparent by ensuring these clusters make dissimilar shapes. 

The MDS layout algorithm tends to depict cycle motifs as (roughly) circles similar to our defined motif, and this is reflected in the data; see \autoref{tab:q5}.
Though the star motif is isomorphic to the cycle, comparing the star shape introduced noise that caused participants to not recognize the same structure, another point to show that it is important to represent similar structures with similar shapes. We note as well the star motif is not a planar drawing, where the cycle is. 
Further motif comparison can be found in the appendix.

\section{Conclusion}
\label{sec:limitations}

To better understand the role of spontaneous processing on the representation of sub-graphs in node-link diagrams, we conducted a user experiment to address research questions relating to the perception of both structure (relations in the sub-graph) and shape (visual form). Amongst other findings, we conclude that presenting identical motifs using identical (or rotated) visual  motif drawings makes a difference to the ease with which they can be matched. 


We considered small motifs (5-8 nodes) in sparse, relatively small graphs. Further work includes determining whether the same results hold for larger motifs or graphs, or for other graph representation idioms (e.g., adjacency matrices or arc diagrams), since the same pattern-matching and bottom-up spontaneous processing issues also apply. Larger, different motifs could be studied, integration into denser graphs, or considering more (and more radical) variations of shape and structure.

Since we have shown that the shape of a motif affects how an individual recognizes it, layout algorithms that extract similar graph motifs and draw them using a similar form would highlight their similarity. Techniques exist to accommodate user constraints such as node positions~\cite{bohringer1990using,kamps1996constraint,yu:PacificVis:2022}, but combining this idea with motif extraction may present algorithmic problems and perceptual outcomes.

\paragraph*{Acknowledgement:} This paper is the result of a collaboration initiated at Dagstuhl Seminar 23051, ``Perception in Network Visualization'', February 2023.

\bibliographystyle{splncs04}
\bibliography{main}

\appendix 

\section{Motif comparison}
\label{sec:motif-compare}
While our overall research aim is not to compare the performance of the particular motifs that we have selected, \autoref{tab:motf-compare} shows the overall accuracy of the data for each motif, firstly focusing on the fixed data (all motifs have a well-formed shape) and then on the flexible data (motifs are distorted).

\begin{table}[htbp]
    \centering
    \bgroup
    \def\arraystretch{1.2}
    \begin{tabular}{|c|c c c c c|}
    \hline
    & bi-clique  & clique  & cycle  & double-cycle  & star \\
    \hline
    All (fixed) & 0.597 & 0.671 & 0.665 & 0.666 & 0.616 \\
    All (flexible) & 0.401 & 0.553 & 0.516 & 0.438 & 0.346\\
    SS (fixed) & 0.543 & 0.861 & 0.792 & 0.712 & 0.866\\ 
      
    SD (fixed) & 0.37 & 0.398 & 0.532 & 0.311 & 0.381\\ 
      
    DS (fixed) & 0.007 & 0.198 & 0.08 & 0.186 & 0.021\\ 
      
    DD (fixed) & 0.313 & 0.401 & 0.308 & 0.256 & 0.375\\ 
      
    SS (flexible) & 0.049 & 0.214 & 0.239 & 0.141 & 0.004\\ 
      
    SD (flexible) & 0.142 & 0.247 & 0.291 & 0.214 & 0.115\\ 
      
    DS (flexible) & 0.129 & 0.198 & 0.047 & 0.192 & 0.067\\ 
      
    DD (flexible) & 0.378 & 0.206 & 0.449 & 0.487 & 0.317\\ 
     \hline
    \end{tabular}
    \egroup
    \vspace{0.25cm}
    \caption{Mean accuracy for both the fixed and flexible drawings for each type of trial.}
    \label{tab:motf-compare}
    \vspace{-1cm}
\end{table}

The bi-clique appears to be the most difficult motif to match – possibly due to the fact that, unlike the other motifs, it only has one axis of symmetry. The star motif under a `flexible’ unconstrained layout has worse accuracy; force-directed layout of course, unravels the structure to some extent, perhaps being represented as a simple cycle (to which it is isomorphic).

\section{Complete List of Stimuli}
\begin{figure}
    \centering
    \scriptsize
    \resizebox{!}{.45\textheight}{
    \begin{tabular}{|c|c|c|c|c|c|}
    \hline
    ~ & \thead{Clique} & \thead{Biclique} & \thead{Cycle} & \thead{Double-\\cycle} & \thead{Star}\\
    \hline
    \thead{\rotatebox[origin=c]{90}{BaseN}} &
    \makecell{\includegraphics[height=0.0625\textheight]{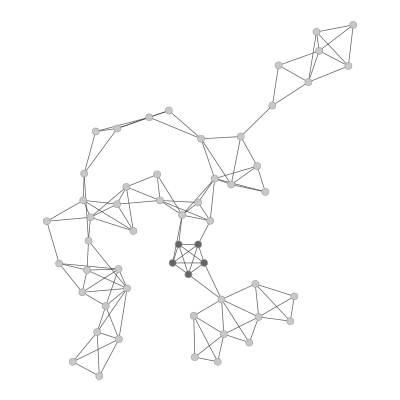}} &
    \makecell{\includegraphics[height=0.0625\textheight]{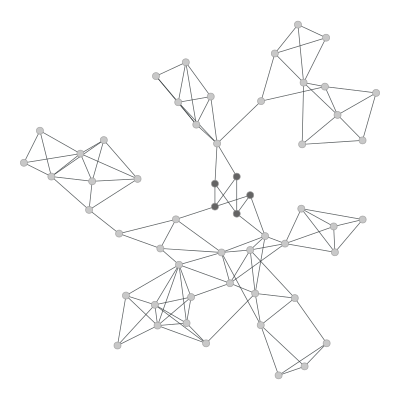}} &
    \makecell{\includegraphics[height=0.0625\textheight]{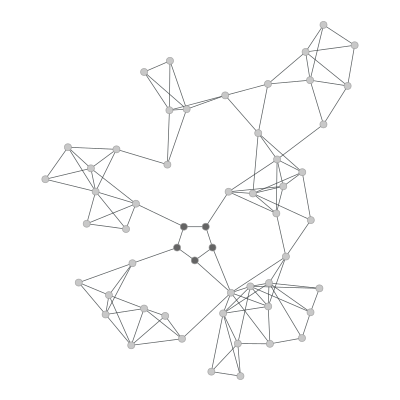}} &
    \makecell{\includegraphics[height=0.0625\textheight]{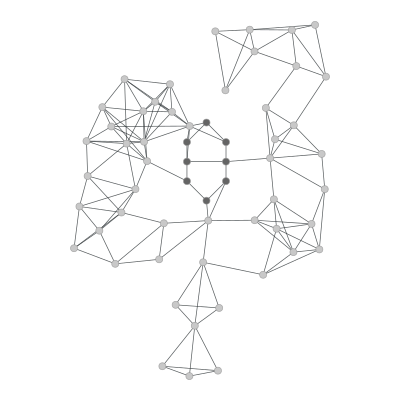}} &
    \makecell{\includegraphics[height=0.0625\textheight]{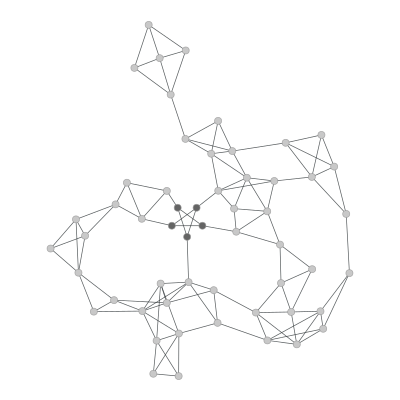}} \\
    \hline
    \thead{\rotatebox[origin=c]{90}{BaseS}} &
    \makecell{\includegraphics[height=0.0625\textheight]{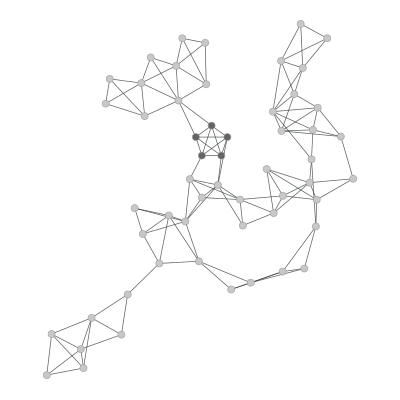}} &
    \makecell{\includegraphics[height=0.0625\textheight]{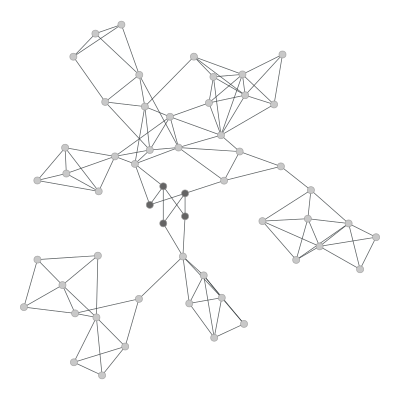}} &
    \makecell{\includegraphics[height=0.0625\textheight]{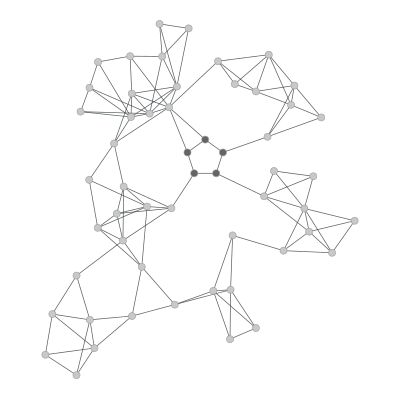}} &
    \makecell{\includegraphics[height=0.0625\textheight]{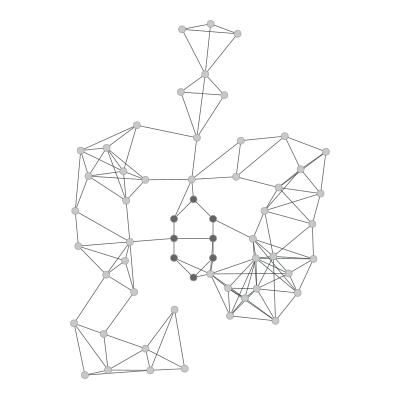}} &
    \makecell{\includegraphics[height=0.0625\textheight]{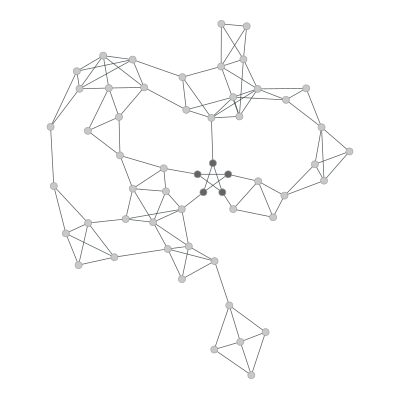}} \\
    \hline
    \thead{\rotatebox[origin=c]{90}{BaseE}} &
    \makecell{\includegraphics[height=0.0625\textheight]{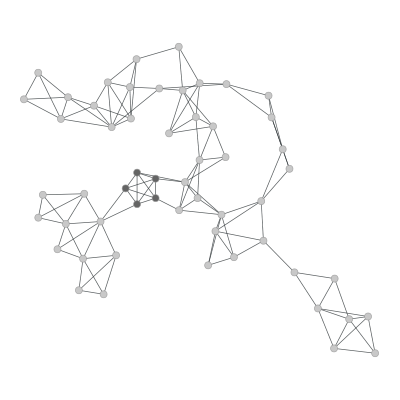}} &
    \makecell{\includegraphics[height=0.0625\textheight]{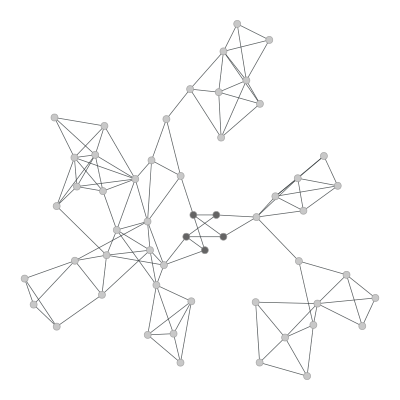}} &
    \makecell{\includegraphics[height=0.0625\textheight]{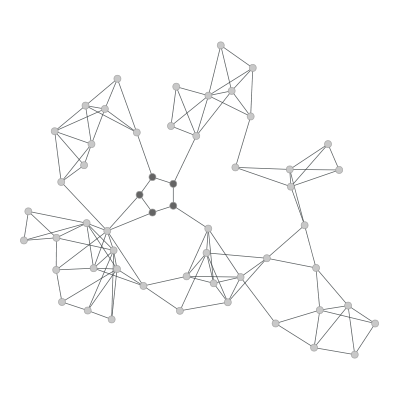}} &
    \makecell{\includegraphics[height=0.0625\textheight]{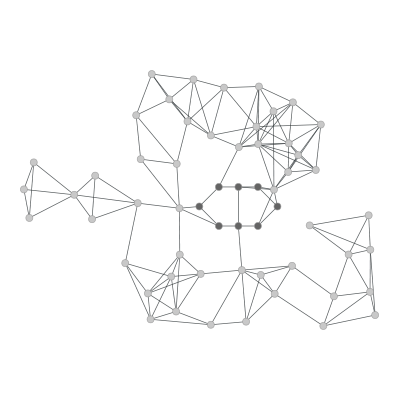}} &
    \makecell{\includegraphics[height=0.0625\textheight]{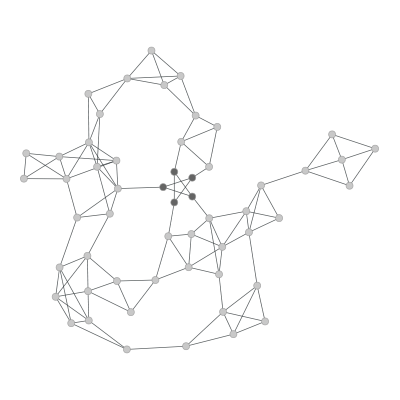}} \\
    \hline
    \thead{\rotatebox[origin=c]{90}{BaseW}} &
    \makecell{\includegraphics[height=0.0625\textheight]{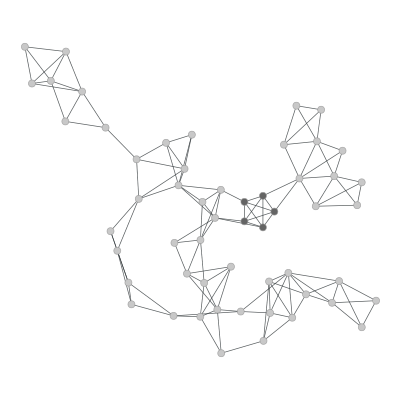}} &
    \makecell{\includegraphics[height=0.0625\textheight]{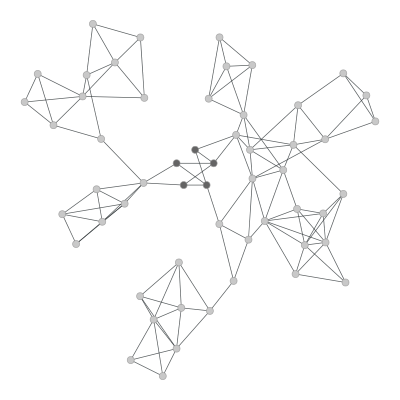}} &
    \makecell{\includegraphics[height=0.0625\textheight]{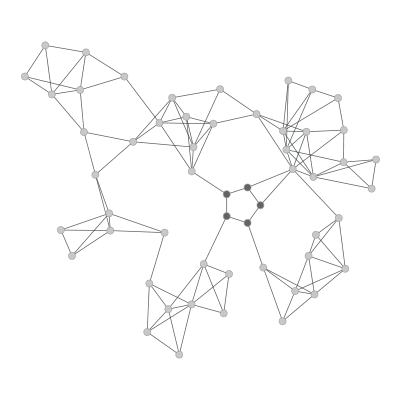}} &
    \makecell{\includegraphics[height=0.0625\textheight]{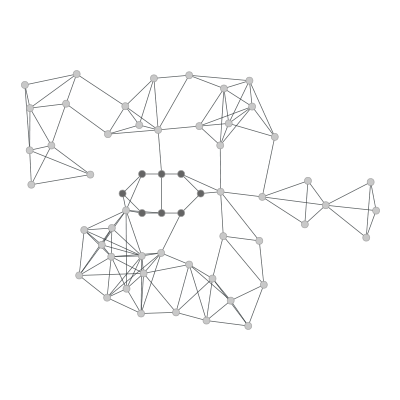}} &
    \makecell{\includegraphics[height=0.0625\textheight]{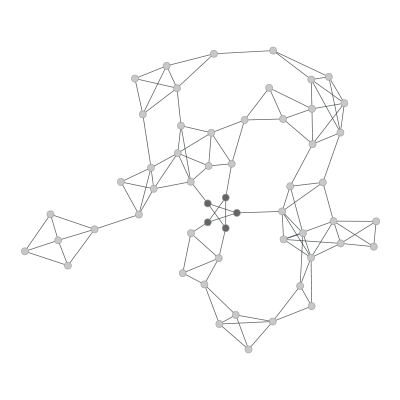}} \\
    \hline
    \thead{\rotatebox[origin=c]{90}{SS1}} & 
    \makecell{\includegraphics[height=0.0625\textheight]{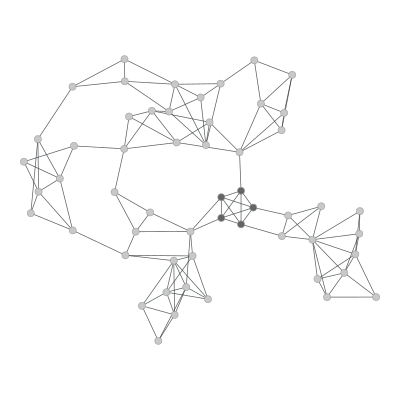}} &
    \makecell{\includegraphics[height=0.0625\textheight]{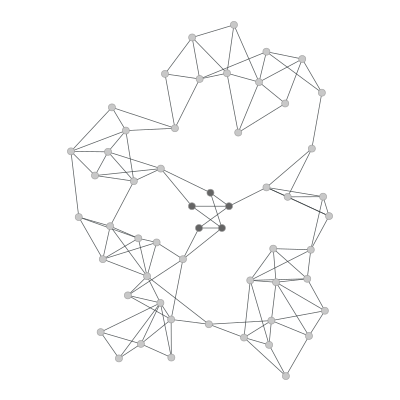}} &
    \makecell{\includegraphics[height=0.0625\textheight]{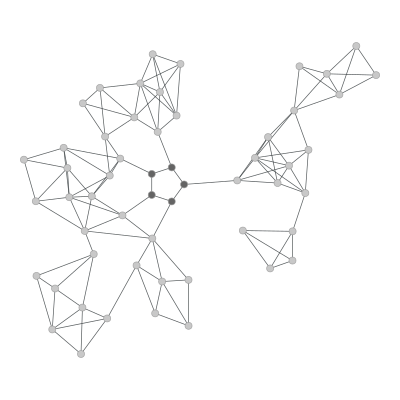}} &
    \makecell{\includegraphics[height=0.0625\textheight]{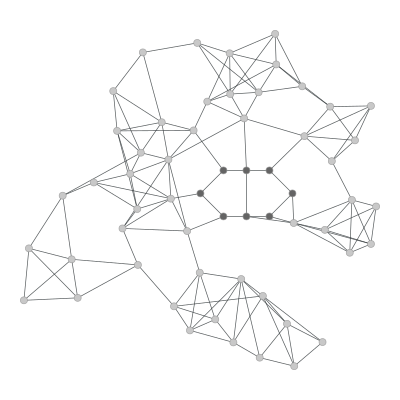}} & 
    \makecell{\includegraphics[height=0.0625\textheight]{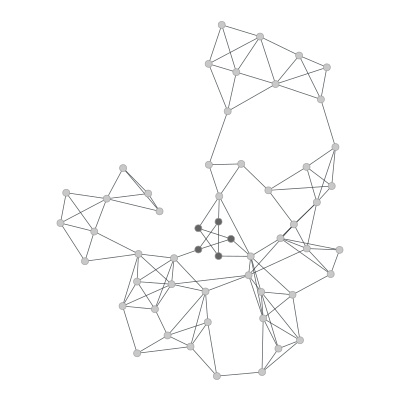}} \\
    \hline
    \thead{\rotatebox[origin=c]{90}{SS2}} & 
    \makecell{\includegraphics[height=0.0625\textheight]{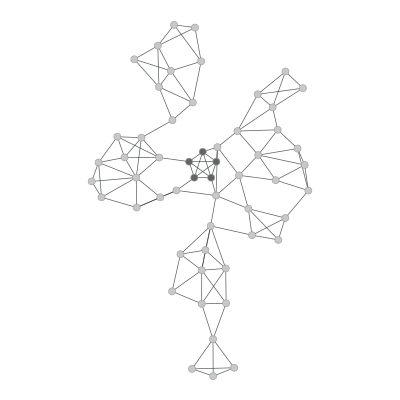}} & 
    \makecell{\includegraphics[height=0.0625\textheight]{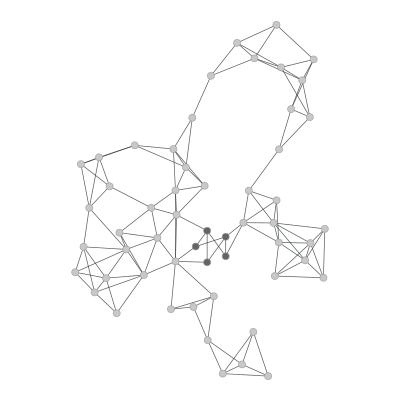}} &
    \makecell{\includegraphics[height=0.0625\textheight]{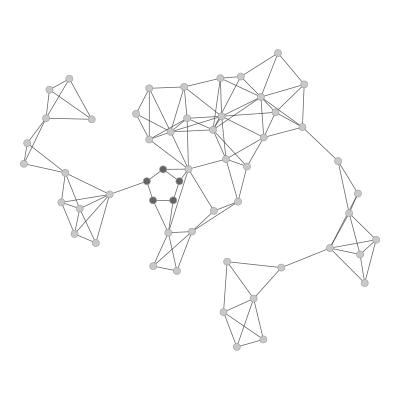}} &
    \makecell{\includegraphics[height=0.0625\textheight]{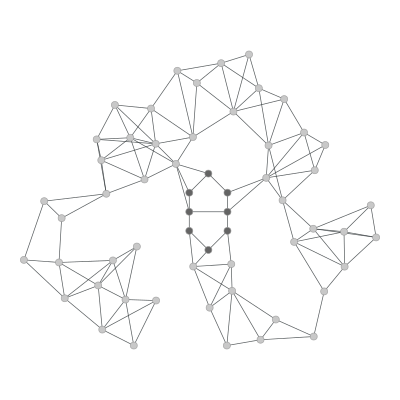}} &
    \makecell{\includegraphics[height=0.0625\textheight]{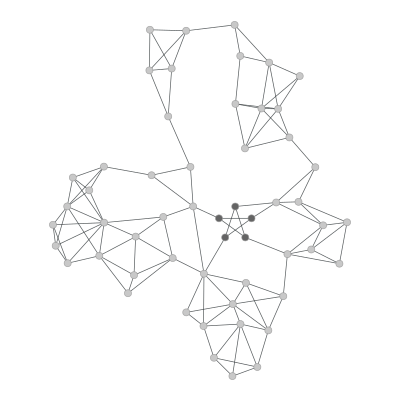}} \\
    \hline
    \thead{\rotatebox[origin=c]{90}{SS3}} &
    \makecell{\includegraphics[height=0.0625\textheight]{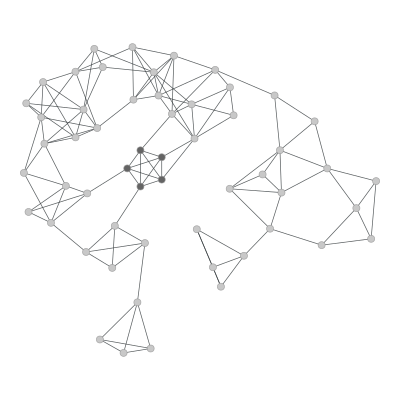}} &
    \makecell{\includegraphics[height=0.0625\textheight]{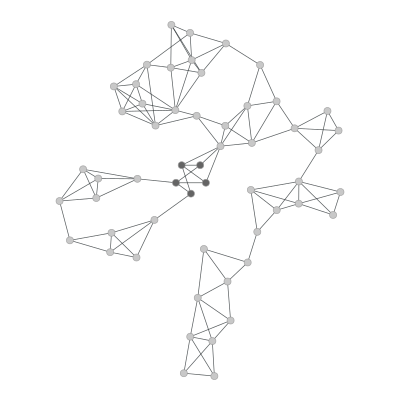}} &
    \makecell{\includegraphics[height=0.0625\textheight]{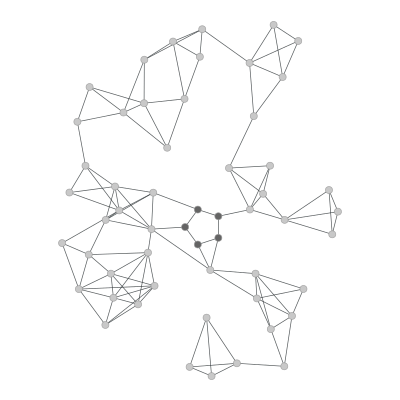}} &
    \makecell{\includegraphics[height=0.0625\textheight]{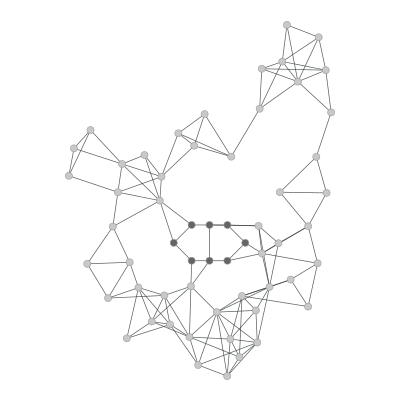}} &
    \makecell{\includegraphics[height=0.0625\textheight]{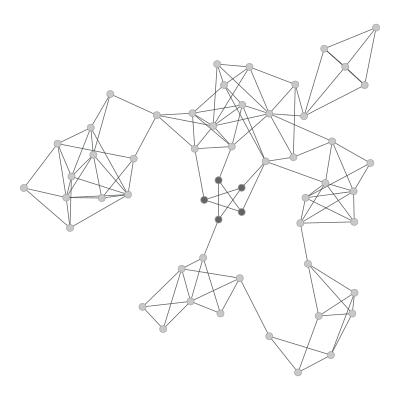}}\\
    \hline
    \thead{\rotatebox[origin=c]{90}{DS1}} & 
    \makecell{\includegraphics[height=0.0625\textheight]{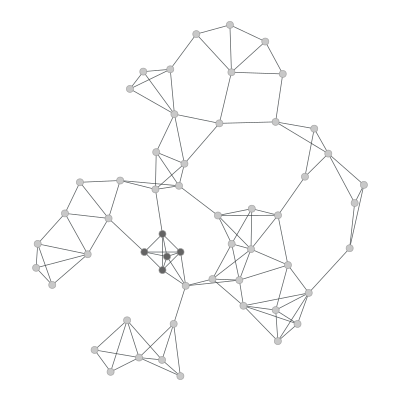}} &
    \makecell{\includegraphics[height=0.0625\textheight]{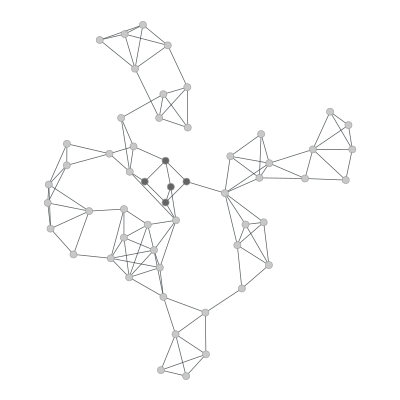}} &
    \makecell{\includegraphics[height=0.0625\textheight]{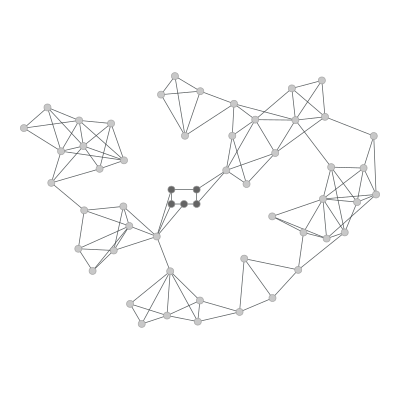}} &
    \makecell{\includegraphics[height=0.0625\textheight]{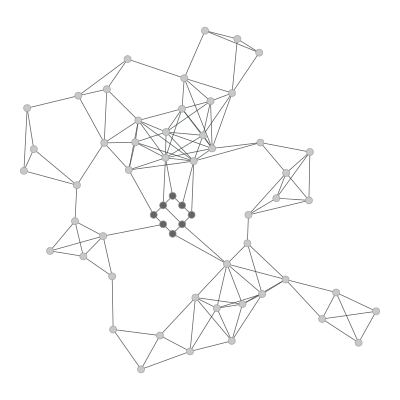}} & 
    \makecell{\includegraphics[height=0.0625\textheight]{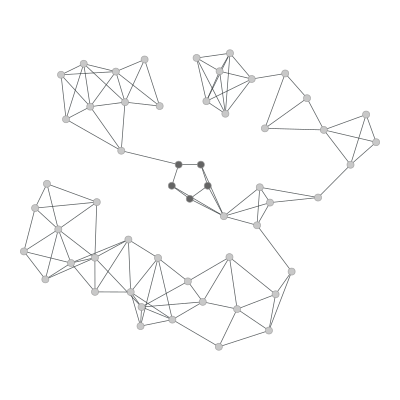}} \\
    \hline
    \thead{\rotatebox[origin=c]{90}{DS2}} & 
    \makecell{\includegraphics[height=0.0625\textheight]{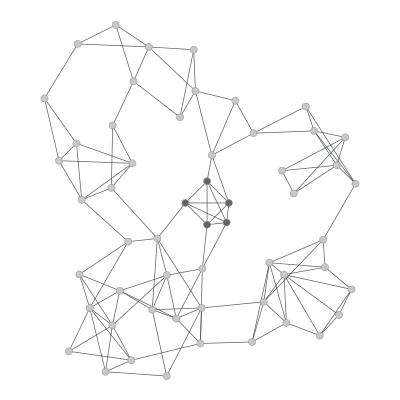}} & 
    \makecell{\includegraphics[height=0.0625\textheight]{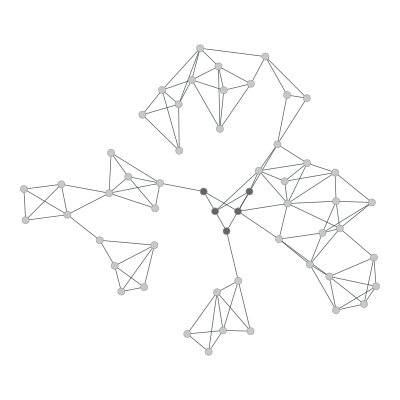}} &
    \makecell{\includegraphics[height=0.0625\textheight]{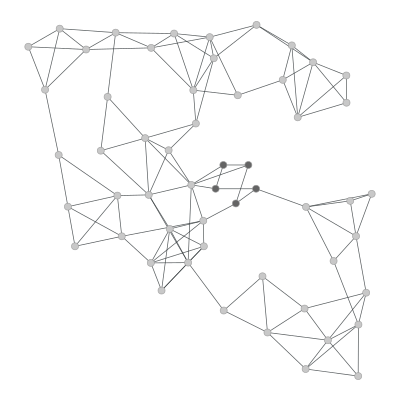}} &
    \makecell{\includegraphics[height=0.0625\textheight]{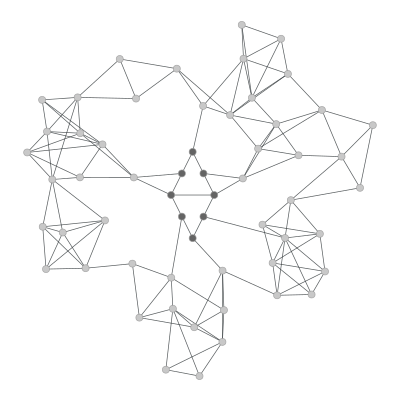}} &
    \makecell{\includegraphics[height=0.0625\textheight]{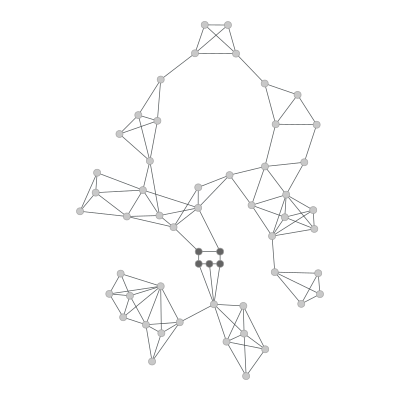}} \\
    \hline
    \thead{\rotatebox[origin=c]{90}{DS3}} &
    \makecell{\includegraphics[height=0.0625\textheight]{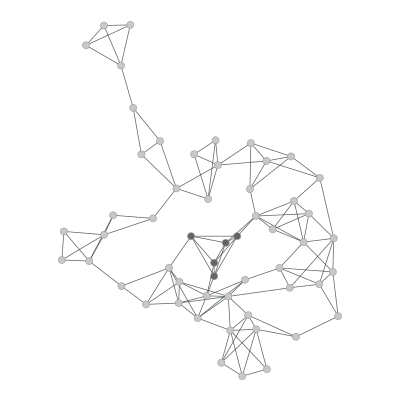}} &
    \makecell{\includegraphics[height=0.0625\textheight]{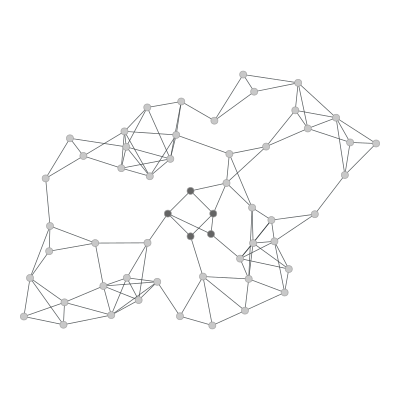}} &
    \makecell{\includegraphics[height=0.0625\textheight]{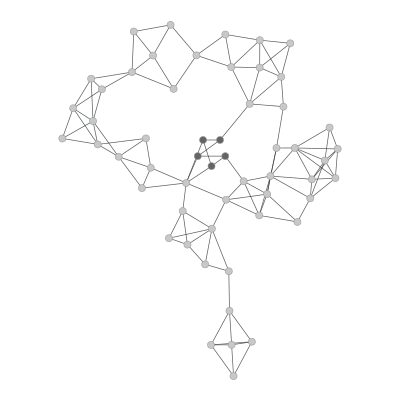}} &
    \makecell{\includegraphics[height=0.0625\textheight]{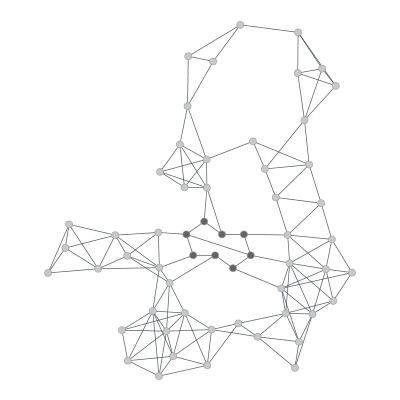}} &
    \makecell{\includegraphics[height=0.0625\textheight]{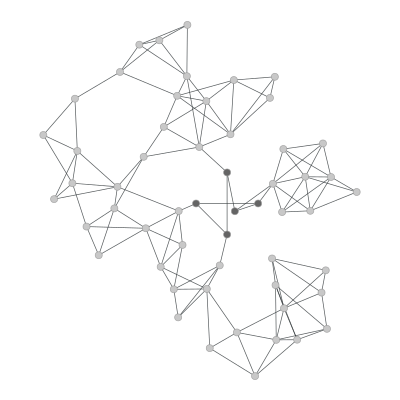}}\\
    \hline
    \thead{\rotatebox[origin=c]{90}{SD1}} & 
    \makecell{\includegraphics[height=0.0625\textheight]{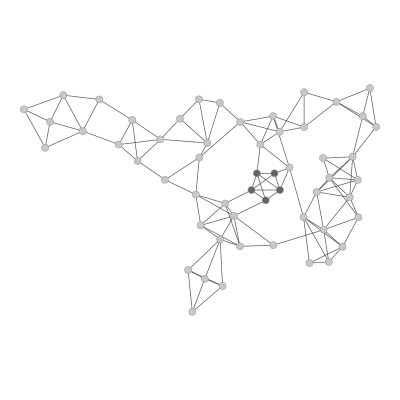}} &
    \makecell{\includegraphics[height=0.0625\textheight]{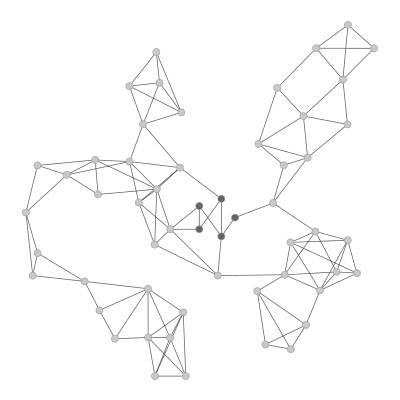}} &
    \makecell{\includegraphics[height=0.0625\textheight]{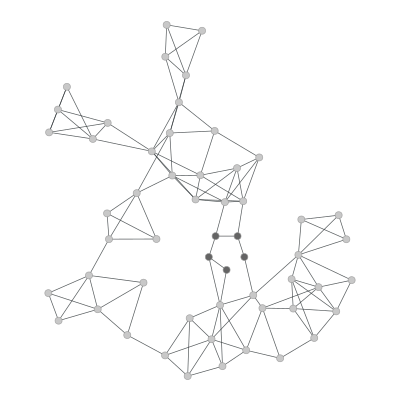}} &
    \makecell{\includegraphics[height=0.0625\textheight]{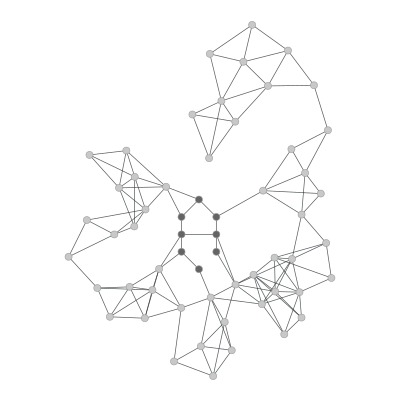}} & 
    \makecell{\includegraphics[height=0.0625\textheight]{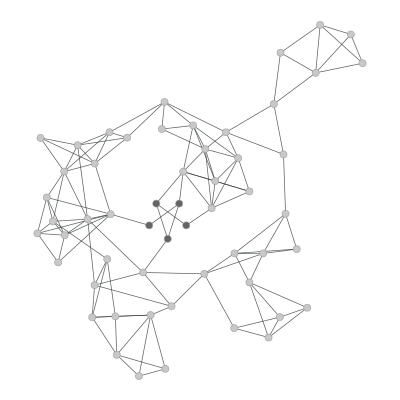}} \\
    \hline
    \thead{\rotatebox[origin=c]{90}{SD2}} & 
    \makecell{\includegraphics[height=0.0625\textheight]{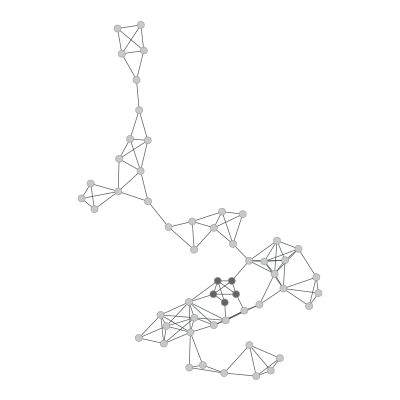}} & 
    \makecell{\includegraphics[height=0.0625\textheight]{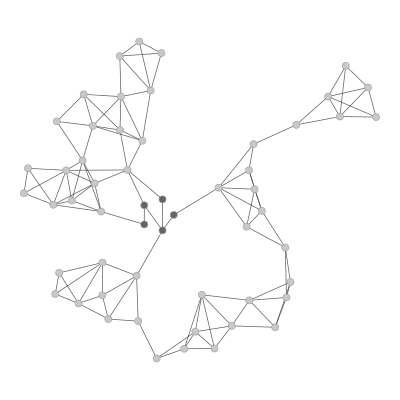}} &
    \makecell{\includegraphics[height=0.0625\textheight]{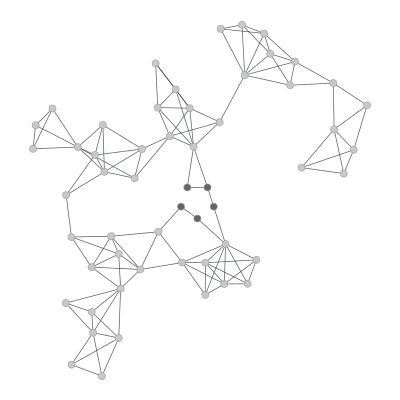}} &
    \makecell{\includegraphics[height=0.0625\textheight]{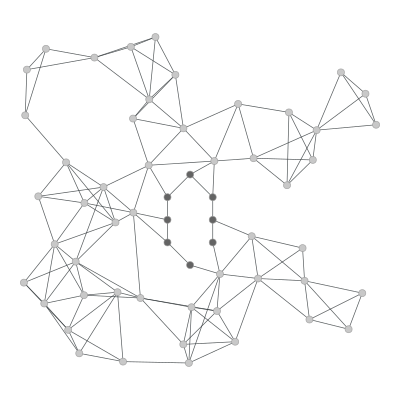}} &
    \makecell{\includegraphics[height=0.0625\textheight]{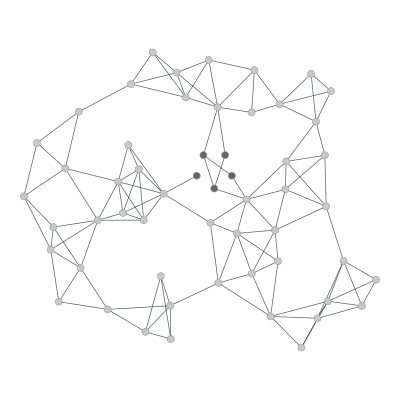}} \\
    \hline
    \thead{\rotatebox[origin=c]{90}{SD3}} &
    \makecell{\includegraphics[height=0.0625\textheight]{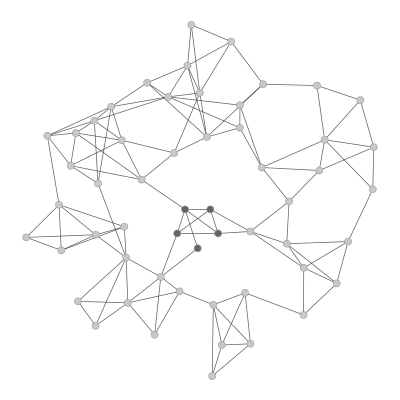}} &
    \makecell{\includegraphics[height=0.0625\textheight]{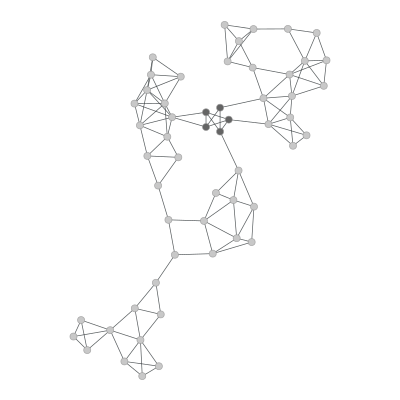}} &
    \makecell{\includegraphics[height=0.0625\textheight]{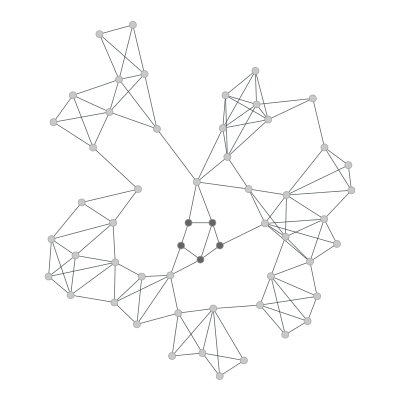}} &
    \makecell{\includegraphics[height=0.0625\textheight]{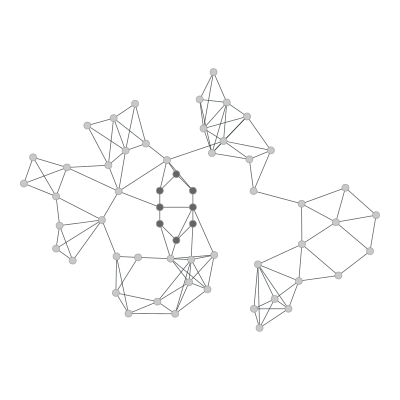}} &
    \makecell{\includegraphics[height=0.0625\textheight]{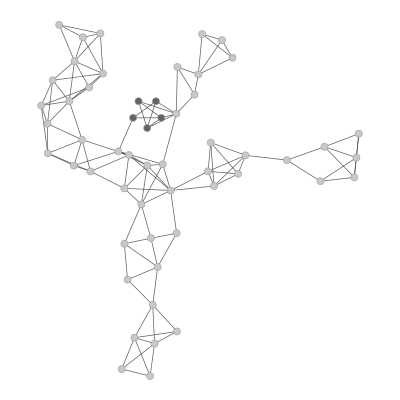}}\\
    \hline
    \thead{\rotatebox[origin=c]{90}{DD1}} & 
    \makecell{\includegraphics[height=0.0625\textheight]{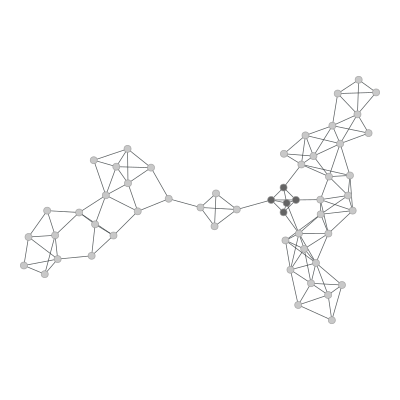}} &
    \makecell{\includegraphics[height=0.0625\textheight]{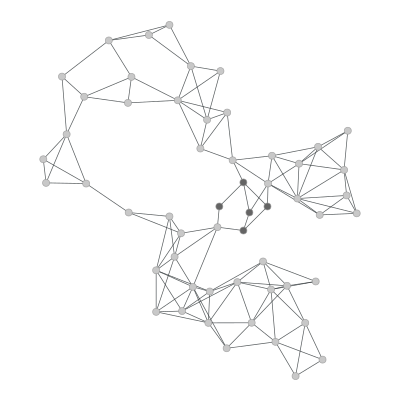}} &
    \makecell{\includegraphics[height=0.0625\textheight]{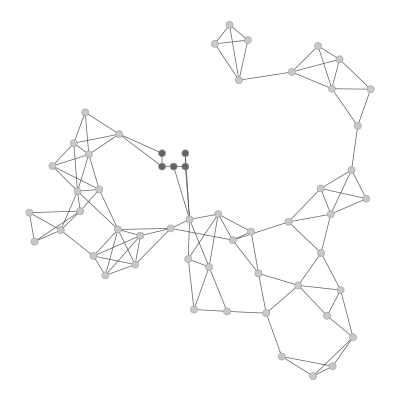}} &
    \makecell{\includegraphics[height=0.0625\textheight]{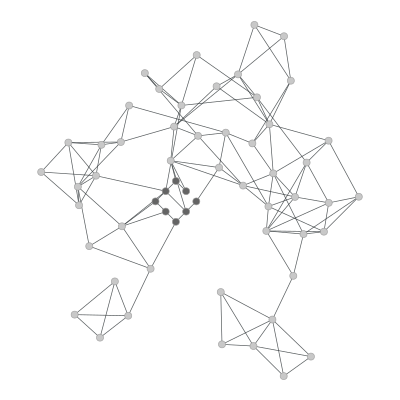}} & 
    \makecell{\includegraphics[height=0.0625\textheight]{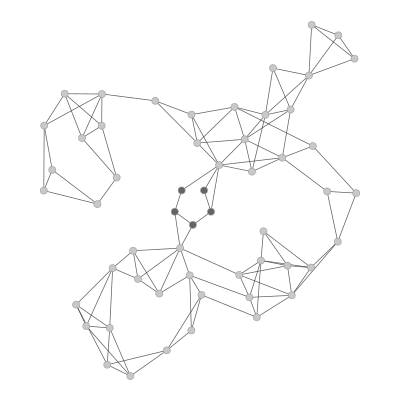}} \\
    \hline
    \thead{\rotatebox[origin=c]{90}{DD2}} & 
    \makecell{\includegraphics[height=0.0625\textheight]{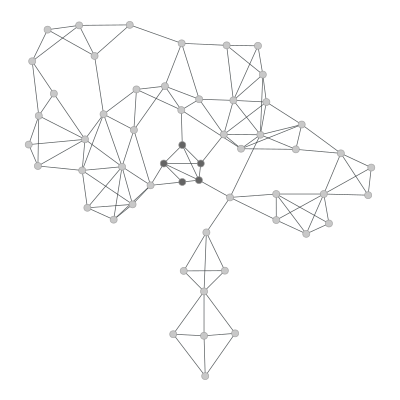}} & 
    \makecell{\includegraphics[height=0.0625\textheight]{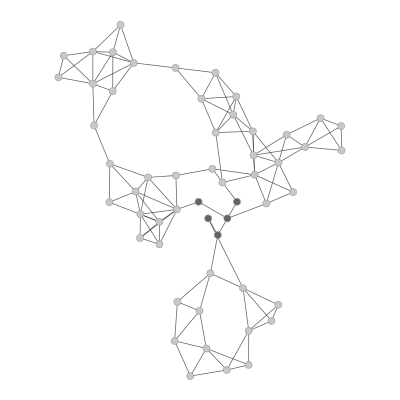}} &
    \makecell{\includegraphics[height=0.0625\textheight]{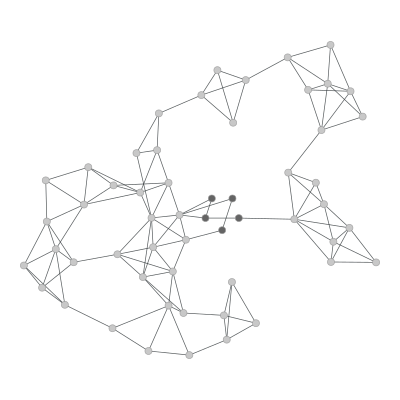}} &
    \makecell{\includegraphics[height=0.0625\textheight]{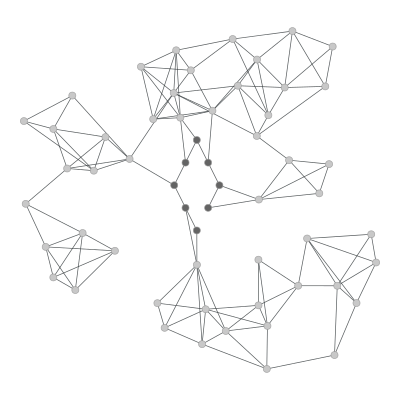}} &
    \makecell{\includegraphics[height=0.0625\textheight]{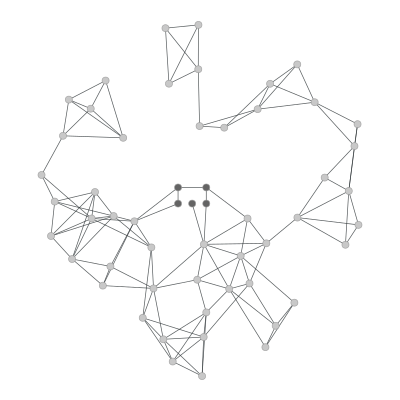}} \\
    \hline
    \thead{\rotatebox[origin=c]{90}{DD3}} &
    \makecell{\includegraphics[height=0.0625\textheight]{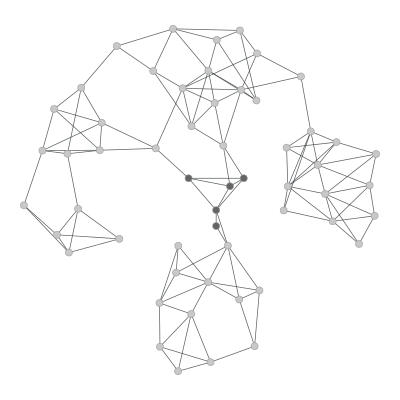}} &
    \makecell{\includegraphics[height=0.0625\textheight]{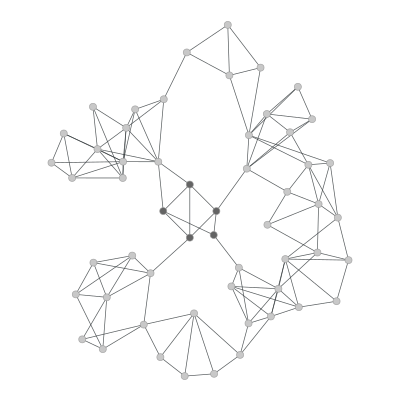}} &
    \makecell{\includegraphics[height=0.0625\textheight]{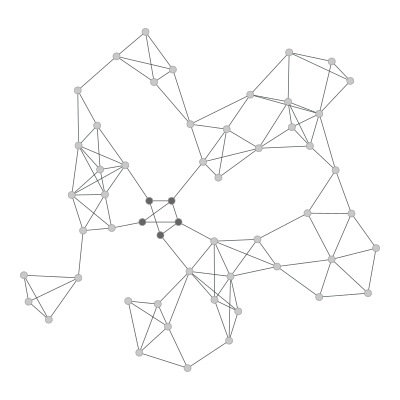}} &
    \makecell{\includegraphics[height=0.0625\textheight]{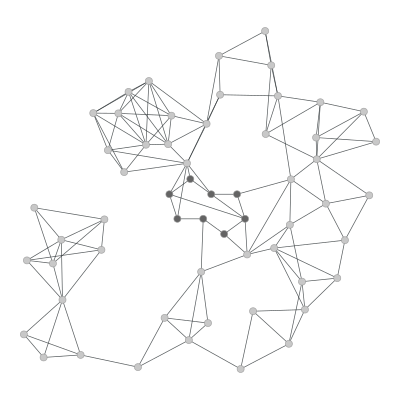}} &
    \makecell{\includegraphics[height=0.0625\textheight]{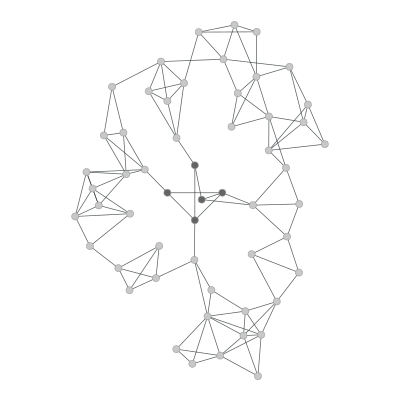}}\\
    \hline
    \end{tabular}
    }    
    \caption{Fixed stimuli used in our experiments}
    \label{fig:static-table}
\end{figure}

\begin{figure}
    \centering
    \scriptsize
    \resizebox{!}{.45\textheight}{
    \begin{tabular}{|c|c|c|c|c|c|}
    \hline
    ~ & \thead{Clique} & \thead{Biclique} & \thead{Cycle} & \thead{Double-\\cycle} & \thead{Star}\\
    \hline
    \thead{\rotatebox[origin=c]{90}{SS1}} & 
    \makecell{\includegraphics[height=0.0625\textheight]{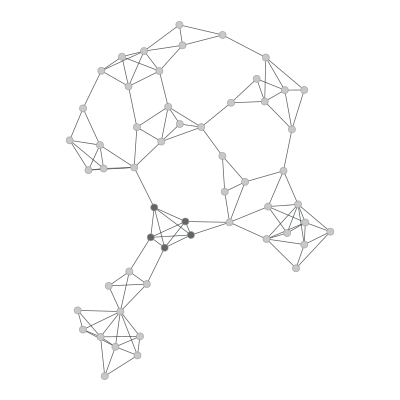}} &
    \makecell{\includegraphics[height=0.0625\textheight]{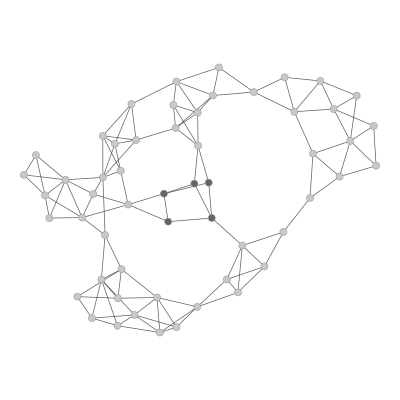}} &
    \makecell{\includegraphics[height=0.0625\textheight]{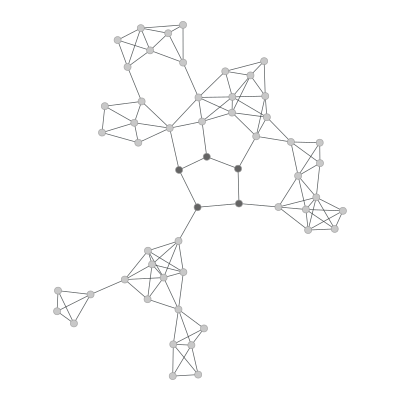}} &
    \makecell{\includegraphics[height=0.0625\textheight]{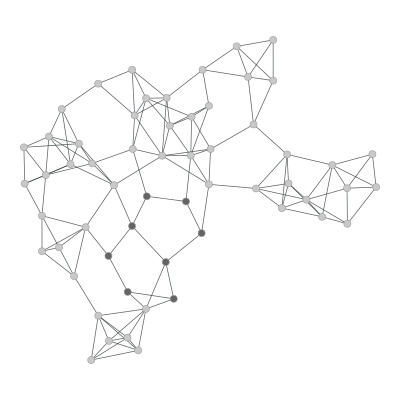}} & 
    \makecell{\includegraphics[height=0.0625\textheight]{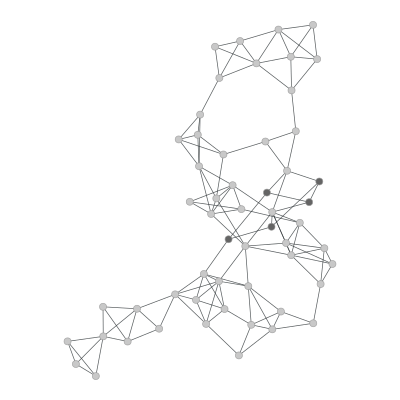}} \\
    \hline
    \thead{\rotatebox[origin=c]{90}{SS2}} & 
    \makecell{\includegraphics[height=0.0625\textheight]{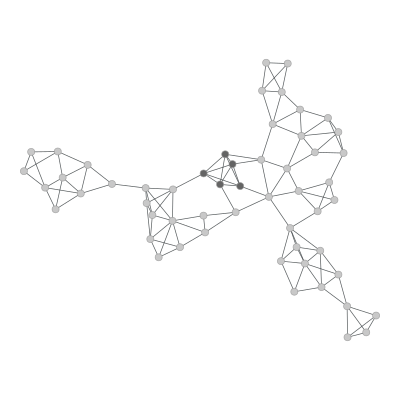}} & 
    \makecell{\includegraphics[height=0.0625\textheight]{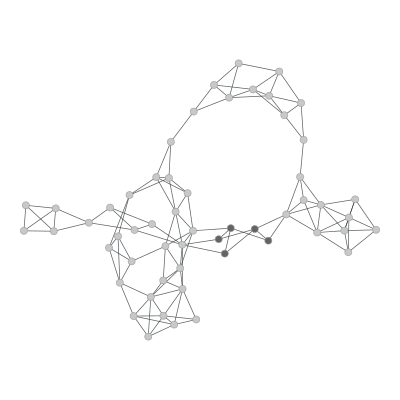}} &
    \makecell{\includegraphics[height=0.0625\textheight]{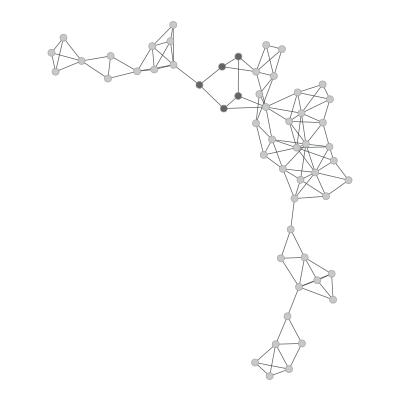}} &
    \makecell{\includegraphics[height=0.0625\textheight]{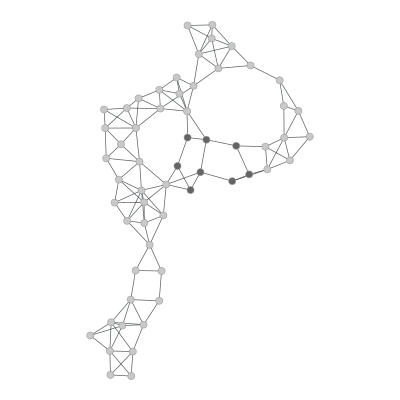}} &
    \makecell{\includegraphics[height=0.0625\textheight]{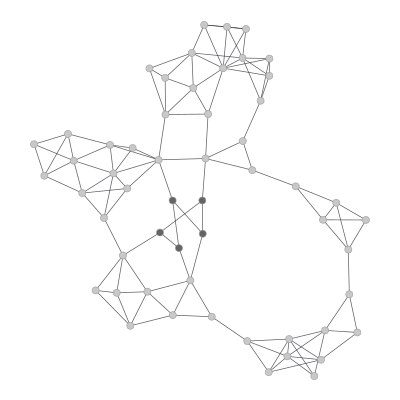}} \\
    \hline
    \thead{\rotatebox[origin=c]{90}{SS3}} &
    \makecell{\includegraphics[height=0.0625\textheight]{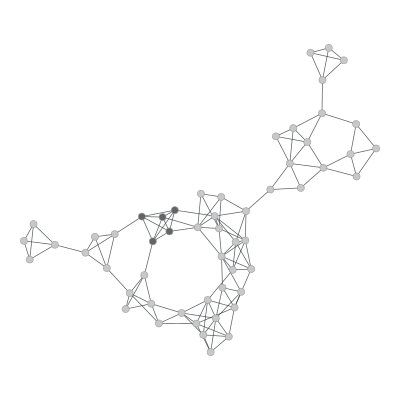}} &
    \makecell{\includegraphics[height=0.0625\textheight]{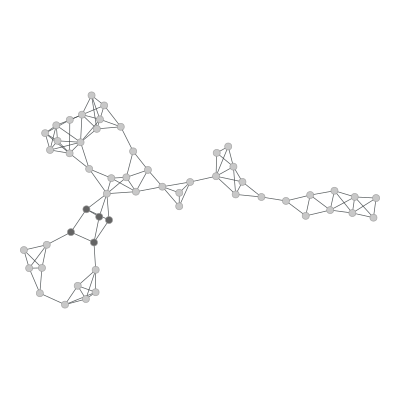}} &
    \makecell{\includegraphics[height=0.0625\textheight]{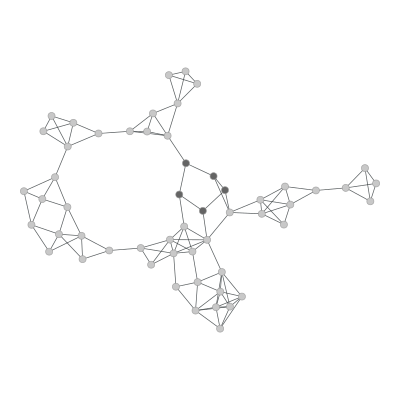}} &
    \makecell{\includegraphics[height=0.0625\textheight]{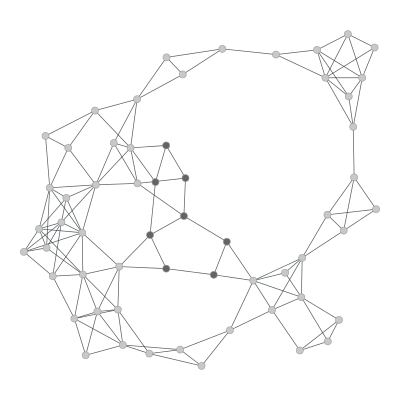}} &
    \makecell{\includegraphics[height=0.0625\textheight]{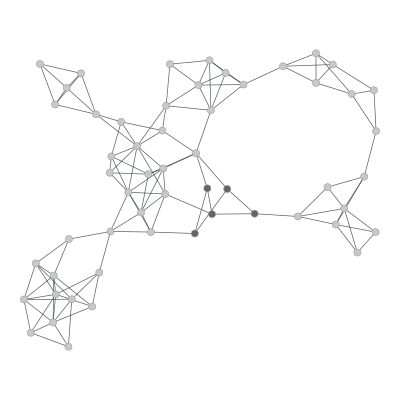}}\\
    \hline
    \thead{\rotatebox[origin=c]{90}{DS1}} & 
    \makecell{\includegraphics[height=0.0625\textheight]{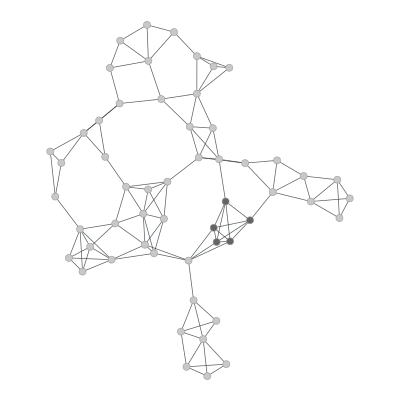}} &
    \makecell{\includegraphics[height=0.0625\textheight]{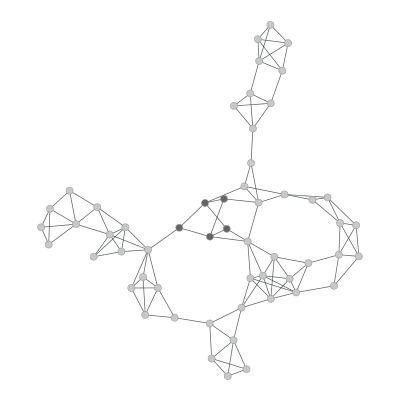}} &
    \makecell{\includegraphics[height=0.0625\textheight]{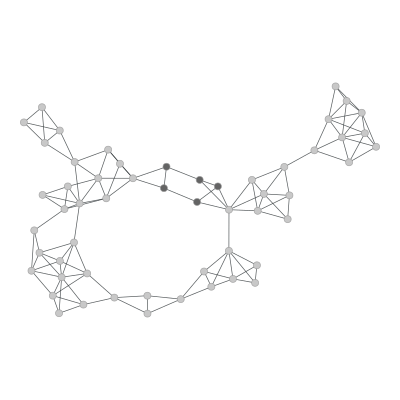}} &
    \makecell{\includegraphics[height=0.0625\textheight]{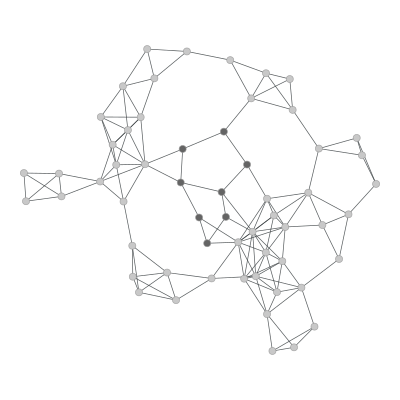}} & 
    \makecell{\includegraphics[height=0.0625\textheight]{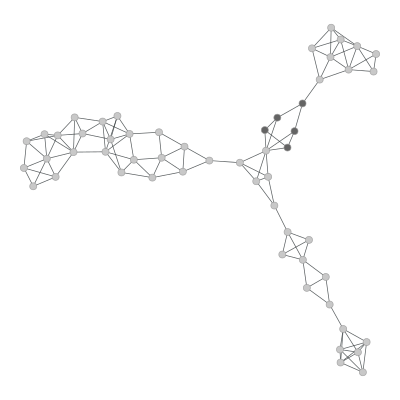}} \\
    \hline
    \thead{\rotatebox[origin=c]{90}{DS2}} & 
    \makecell{\includegraphics[height=0.0625\textheight]{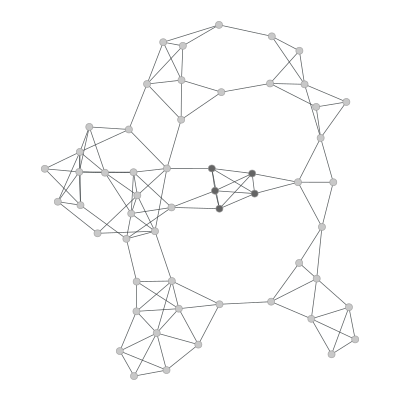}} & 
    \makecell{\includegraphics[height=0.0625\textheight]{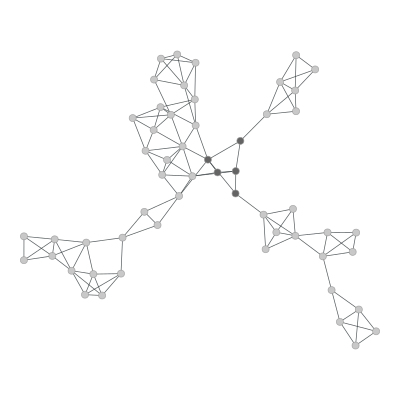}} &
    \makecell{\includegraphics[height=0.0625\textheight]{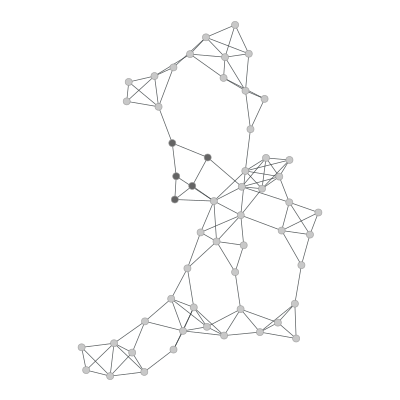}} &
    \makecell{\includegraphics[height=0.0625\textheight]{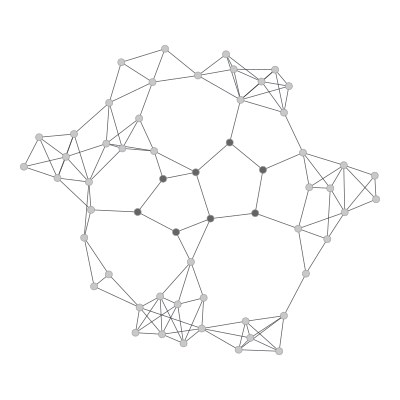}} &
    \makecell{\includegraphics[height=0.0625\textheight]{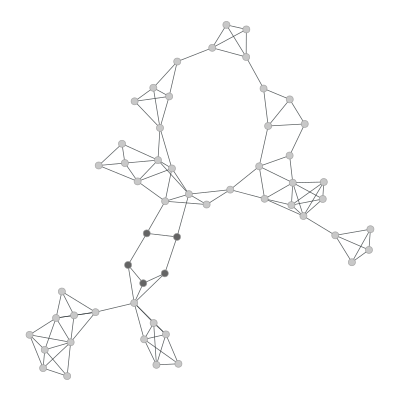}} \\
    \hline
    \thead{\rotatebox[origin=c]{90}{DS3}} &
    \makecell{\includegraphics[height=0.0625\textheight]{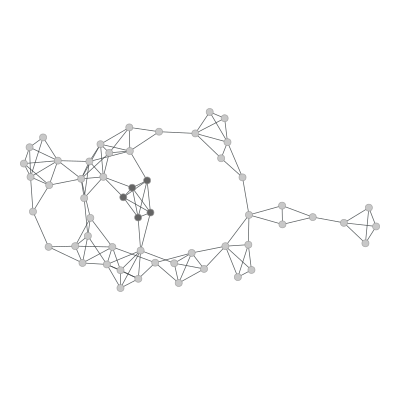}} &
    \makecell{\includegraphics[height=0.0625\textheight]{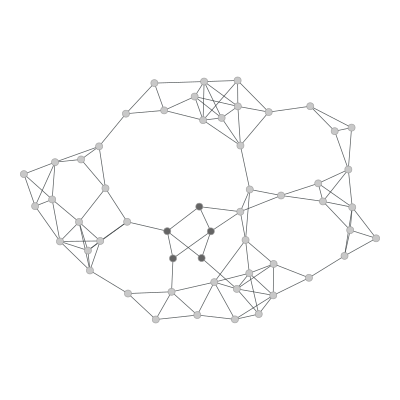}} &
    \makecell{\includegraphics[height=0.0625\textheight]{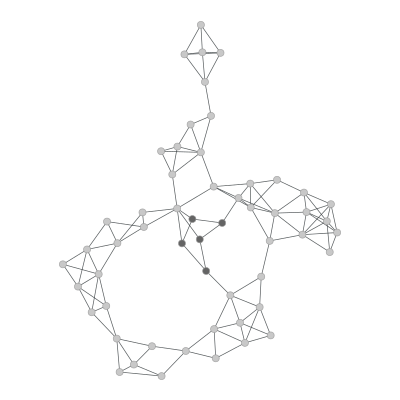}} &
    \makecell{\includegraphics[height=0.0625\textheight]{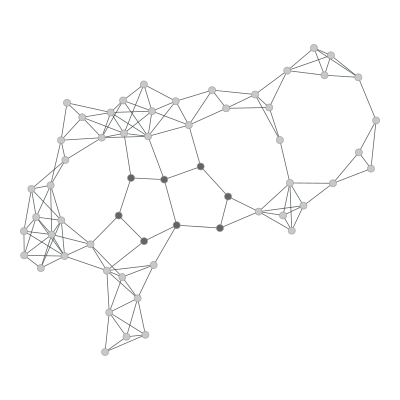}} &
    \makecell{\includegraphics[height=0.0625\textheight]{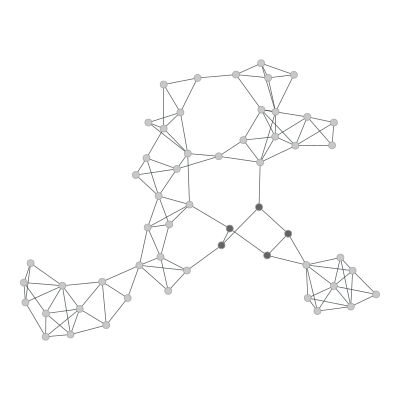}}\\
    \hline
    \thead{\rotatebox[origin=c]{90}{SD1}} & 
    \makecell{\includegraphics[height=0.0625\textheight]{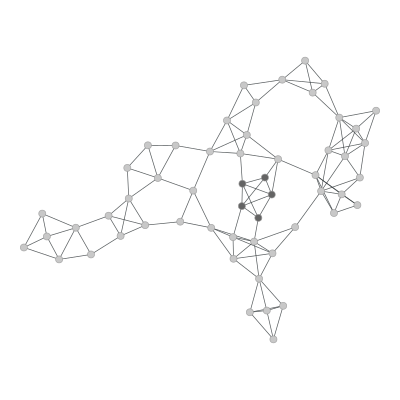}} &
    \makecell{\includegraphics[height=0.0625\textheight]{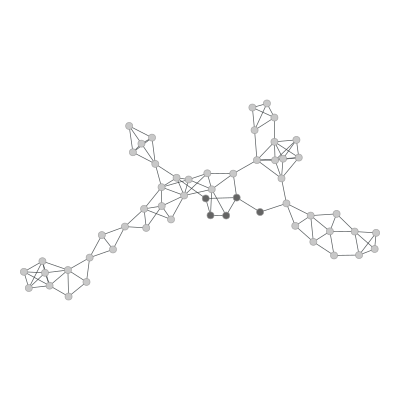}} &
    \makecell{\includegraphics[height=0.0625\textheight]{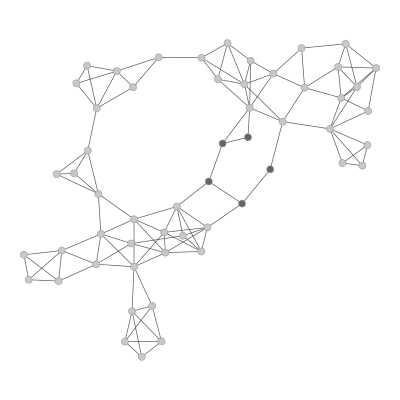}} &
    \makecell{\includegraphics[height=0.0625\textheight]{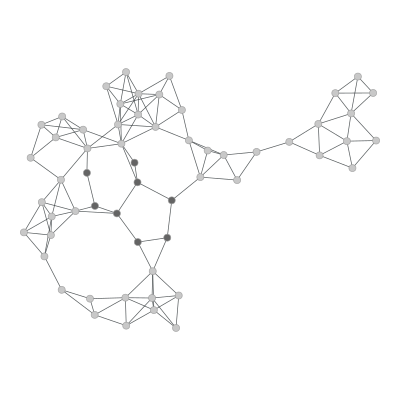}} & 
    \makecell{\includegraphics[height=0.0625\textheight]{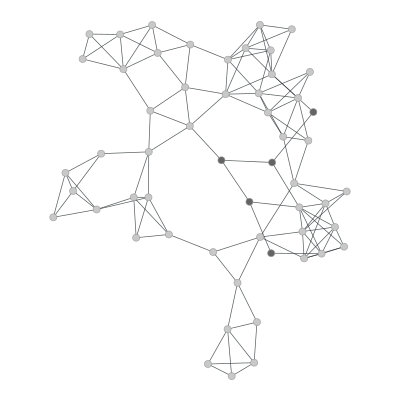}} \\
    \hline
    \thead{\rotatebox[origin=c]{90}{SD2}} & 
    \makecell{\includegraphics[height=0.0625\textheight]{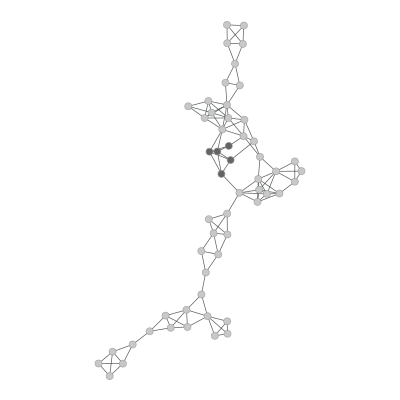}} & 
    \makecell{\includegraphics[height=0.0625\textheight]{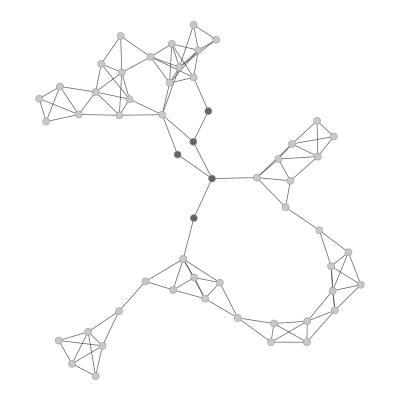}} &
    \makecell{\includegraphics[height=0.0625\textheight]{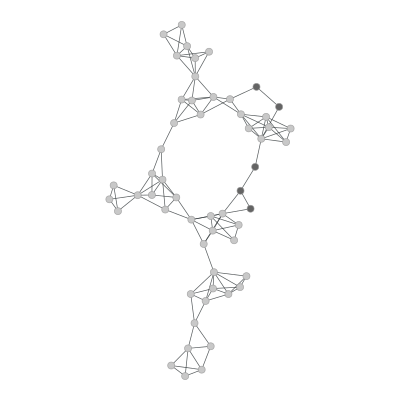}} &
    \makecell{\includegraphics[height=0.0625\textheight]{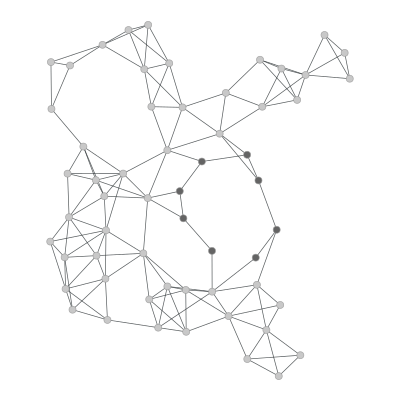}} &
    \makecell{\includegraphics[height=0.0625\textheight]{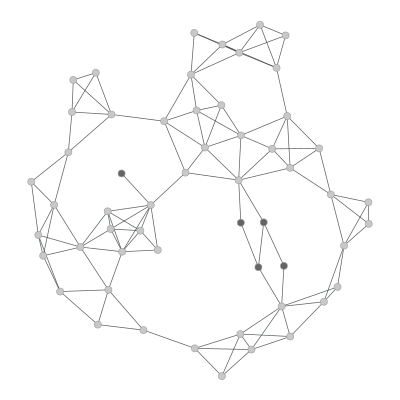}} \\
    \hline
    \thead{\rotatebox[origin=c]{90}{SD3}} &
    \makecell{\includegraphics[height=0.0625\textheight]{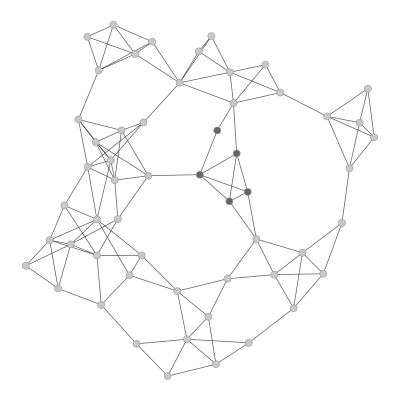}} &
    \makecell{\includegraphics[height=0.0625\textheight]{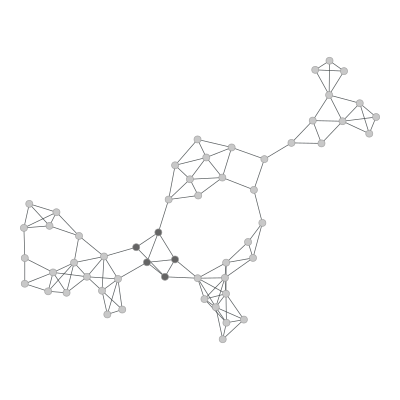}} &
    \makecell{\includegraphics[height=0.0625\textheight]{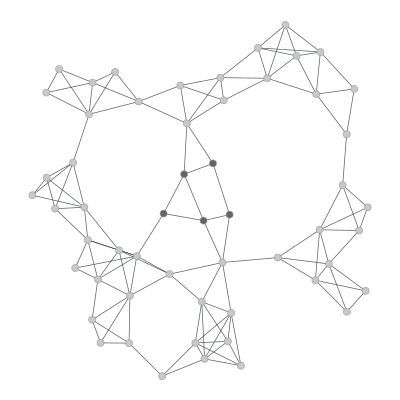}} &
    \makecell{\includegraphics[height=0.0625\textheight]{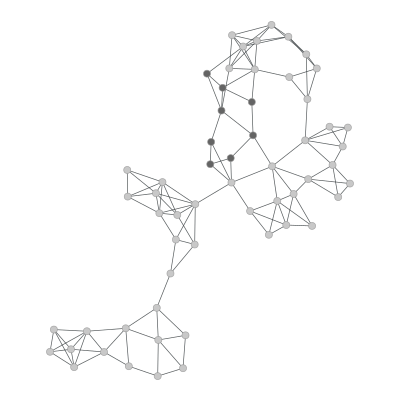}} &
    \makecell{\includegraphics[height=0.0625\textheight]{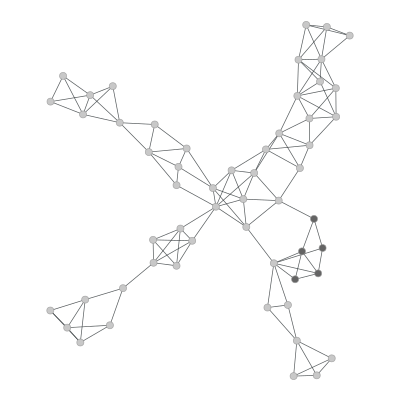}}\\
    \hline
    \thead{\rotatebox[origin=c]{90}{DD1}} & 
    \makecell{\includegraphics[height=0.0625\textheight]{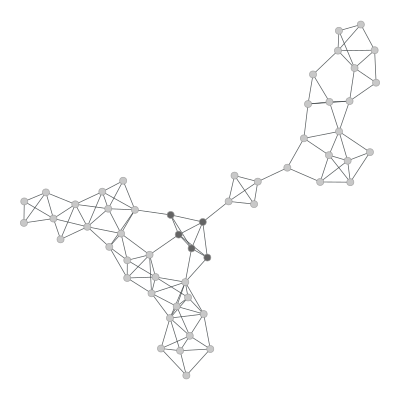}} &
    \makecell{\includegraphics[height=0.0625\textheight]{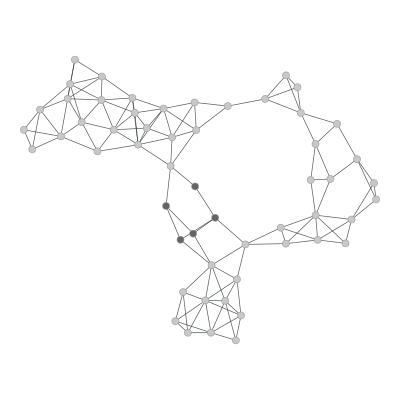}} &
    \makecell{\includegraphics[height=0.0625\textheight]{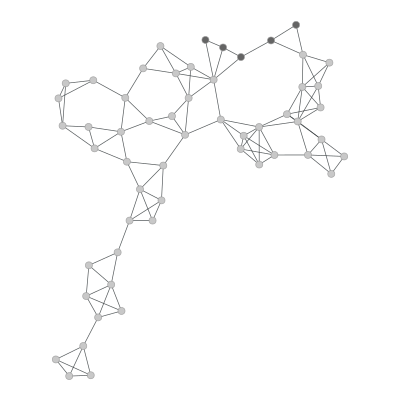}} &
    \makecell{\includegraphics[height=0.0625\textheight]{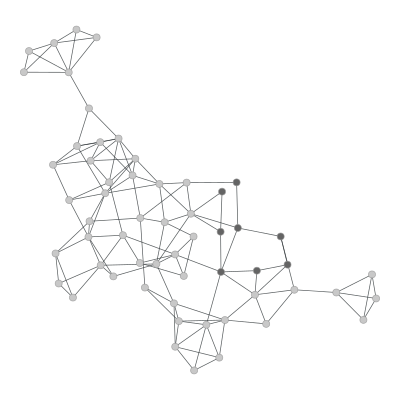}} & 
    \makecell{\includegraphics[height=0.0625\textheight]{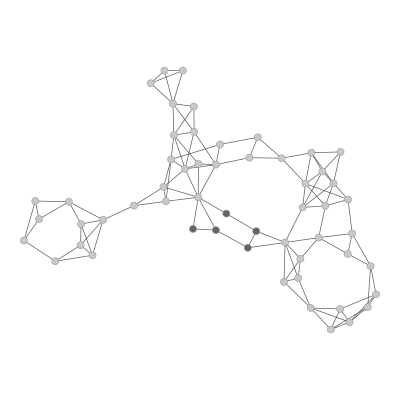}} \\
    \hline
    \thead{\rotatebox[origin=c]{90}{DD2}} & 
    \makecell{\includegraphics[height=0.0625\textheight]{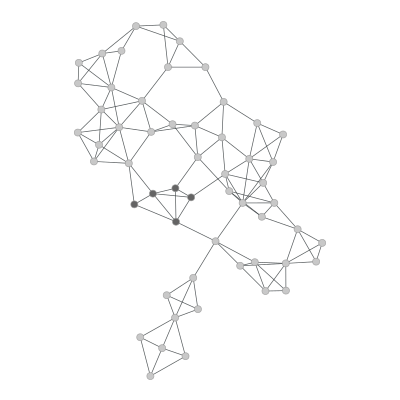}} & 
    \makecell{\includegraphics[height=0.0625\textheight]{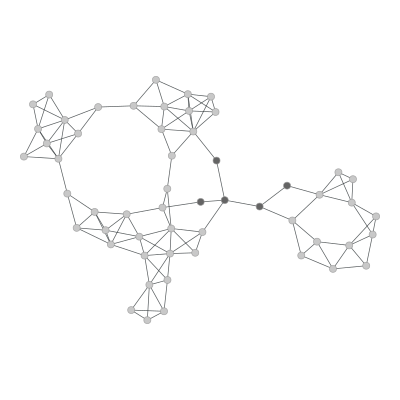}} &
    \makecell{\includegraphics[height=0.0625\textheight]{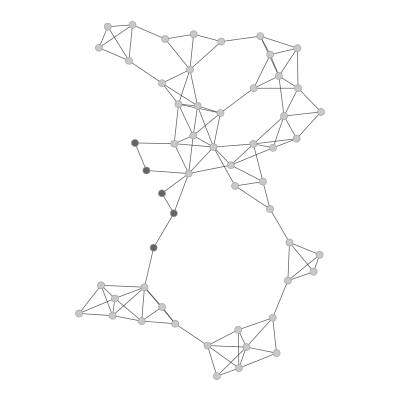}} &
    \makecell{\includegraphics[height=0.0625\textheight]{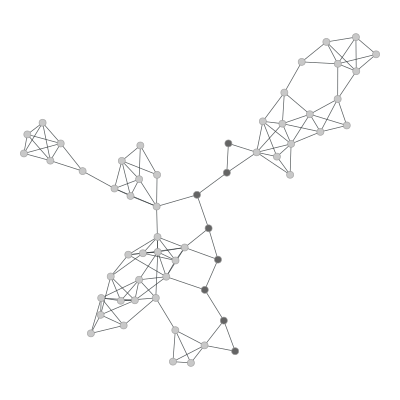}} &
    \makecell{\includegraphics[height=0.0625\textheight]{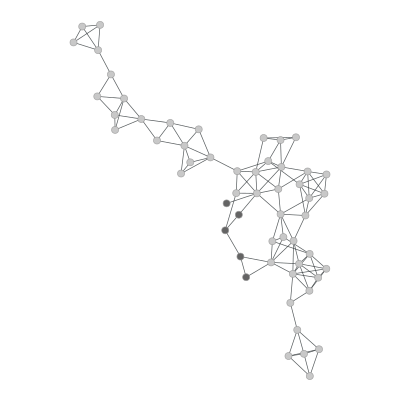}} \\
    \hline
    \thead{\rotatebox[origin=c]{90}{DD3}} &
    \makecell{\includegraphics[height=0.0625\textheight]{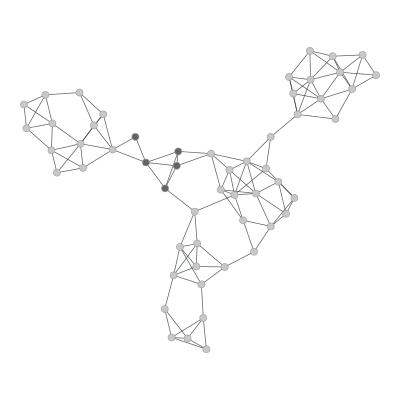}} &
    \makecell{\includegraphics[height=0.0625\textheight]{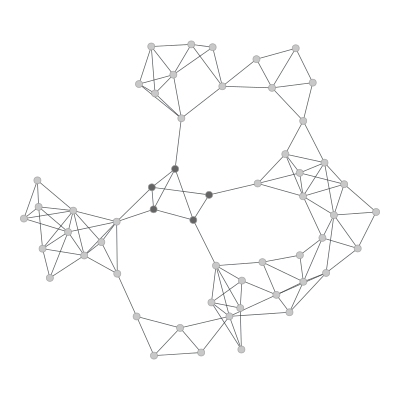}} &
    \makecell{\includegraphics[height=0.0625\textheight]{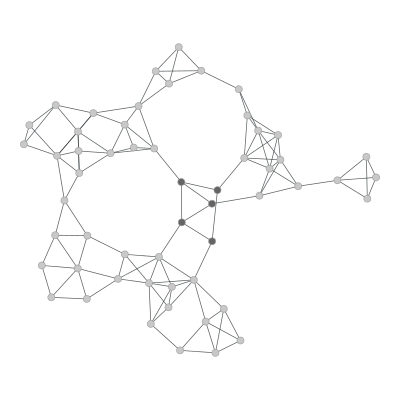}} &
    \makecell{\includegraphics[height=0.0625\textheight]{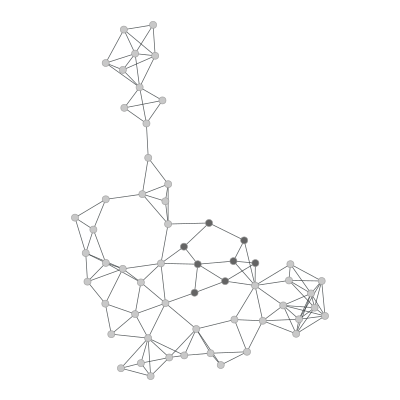}} &
    \makecell{\includegraphics[height=0.0625\textheight]{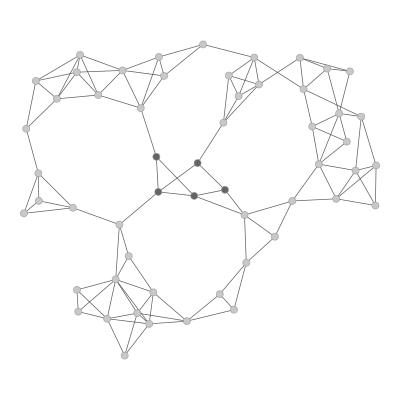}}\\
    \hline
    \thead{\rotatebox[origin=c]{90}{Base}} &
    \makecell{\includegraphics[height=0.0625\textheight]{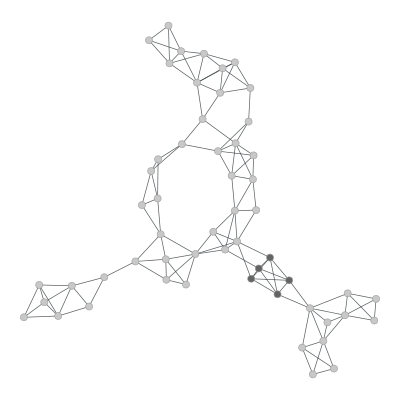}} &
    \makecell{\includegraphics[height=0.0625\textheight]{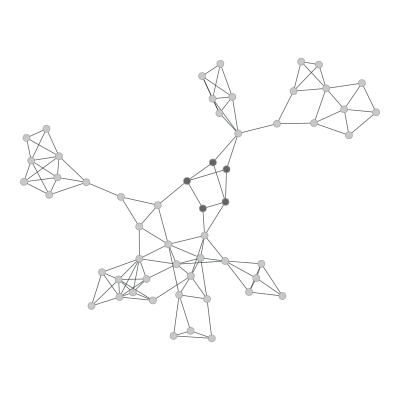}} &
    \makecell{\includegraphics[height=0.0625\textheight]{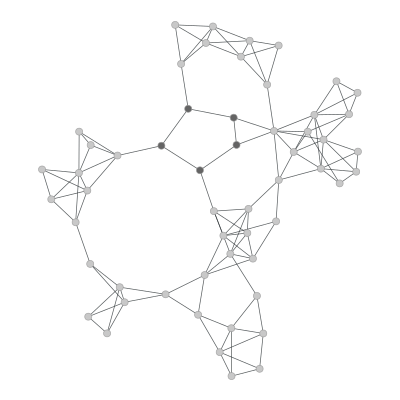}} &
    \makecell{\includegraphics[height=0.0625\textheight]{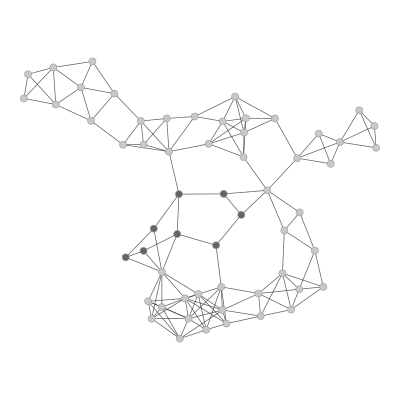}} &
    \makecell{\includegraphics[height=0.0625\textheight]{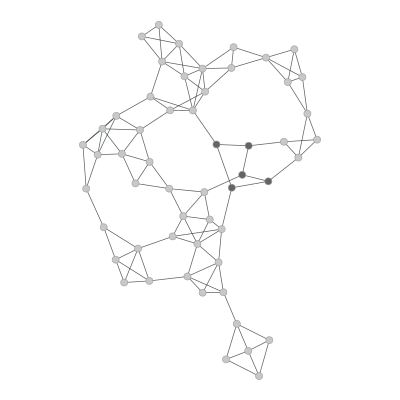}} \\
    \hline
    \end{tabular}
    }
    \caption{Flexible stimuli used in our experiments}
    \label{fig:flex-table}
\end{figure}

A table of all the stimuli used in our experiment is found below. \autoref{fig:static-table} contains all the fixed stimuli and \autoref{fig:flex-table} contains all the flexible stimuli.




\renewcommand\theadalign{bc}
\renewcommand\theadfont{\bfseries}
\renewcommand\theadgape{\Gape[1pt]}
\renewcommand\cellgape{\Gape[0pt]}

\end{document}